\documentclass[]{ecai} 

\usepackage{latexsym}
\usepackage{amssymb}
\usepackage{amsmath}
\usepackage{amsthm}
\usepackage{booktabs}
\usepackage{enumitem}
\usepackage{graphicx}
\usepackage{color}

\usepackage{url}
\usepackage{stmaryrd}
\usepackage{algorithm}
\usepackage{algpseudocode}
\usepackage{dsfont}
\usepackage{adjustbox}
\usepackage{bm}
\usepackage{mathtools}
\newtheorem{theorem}{Theorem}

\newtheorem{definition}{Definition}

\newenvironment{proofsketch}{%
  \proof
}{\endproof}

\newcommand{\tin}{\text{in}}
\newcommand{\tout}{\text{out}}
\newcommand{\ASR}{\text{ASR}}
\newcommand{\LIRA}{\text{LiRA}}
\newcommand{\SMI}{\text{\normalfont SMI}}
\newcommand{\PSMI}{\text{\normalfont PSMI}}
\newcommand{\MI}{\text{\normalfont I}}

\newcommand{\Exp}[1]{\underset{#1}{\mathbb{E}}}
\newcommand{\EE}{\mathbb{E}}
\newcommand{\RR}{\mathbb{R}}
\newcommand{\given}{\ | \ }
\newcommand{\diff}{\mathrm{d}}
\newcommand{\TPR}{\text{\normalfont TPR}}
\newcommand{\FPR}{\text{\normalfont FPR}}
\newcommand{\PVL}{\text{\normalfont PVL}}
\newcommand{\PRC}{\text{\normalfont PRC}}
\newcommand{\PP}{\text{\normalfont P}}
\newcommand{\NN}{\text{\normalfont N}}
\newcommand{\TP}{\text{\normalfont TP}}
\newcommand{\FP}{\text{\normalfont FP}}

\makeatletter
\if@double@blind@
    \newcommand{\safelink}[2]{[Double-blind external link placeholder: #2]}
\else
    \newcommand{\safelink}[2]{\url{#1}}
\fi
\makeatother

\newcommand{\hrho}{\bm \rho}
\newcommand{\hlambda}{\bm \lambda}
\newcommand{\hmu}{\bm \mu}
\newcommand{\htau}{\bm \tau}
\newcommand{\heta}{\bm \eta}

\usepackage{amsmath,amsfonts,bm}

\def\eqref#1{equation~\ref{#1}}

\def\1{\bm{1}}

\def\rmX{{\mathbf{X}}}

\DeclareMathAlphabet{\mathsfit}{\encodingdefault}{\sfdefault}{m}{sl}
\SetMathAlphabet{\mathsfit}{bold}{\encodingdefault}{\sfdefault}{bx}{n}

\begin{document}


\begin{frontmatter}


\paperid{3664} 


\title{Predicting memorization within Large Language Models fine-tuned for classification}

\author[A]{\fnms{Jérémie}~\snm{Dentan}\orcid{0009-0001-5561-8030}\thanks{Corresponding Author. Email: jeremie.dentan@polytechnique.edu.\\See code at: \safelink{https://github.com/orailix/predict_llm_memorization}{GitHub}\\This paper has been accepted for publication at ECAI 2025.}}
\author[A,B]{\fnms{Davide}~\snm{Buscaldi}\orcid{0000-0003-1112-3789}}
\author[C]{\fnms{Aymen}~\snm{Shabou}\orcid{0000-0001-8933-7053}} 
\author[A]{\fnms{Sonia}~\snm{Vanier}\orcid{0000-0001-6390-8882}} 

\address[A]{LIX (École Polytechnique, IP Paris, CNRS)}
\address[B]{LIPN (Université Sorbonne Paris Nord)}
\address[C]{Crédit Agricole SA}


\begin{abstract}
Large Language Models have received significant attention due to their abilities to solve a wide range of complex tasks. However these models memorize a significant proportion of their training data, posing a serious threat when disclosed at inference time. To mitigate this unintended memorization, it is crucial to understand what elements are memorized and why. This area of research is largely unexplored, with most existing works providing \textit{a posteriori} explanations. To address this gap, we propose a new approach to detect memorized samples \textit{a priori} in LLMs fine-tuned for classification tasks. This method is effective from the early stages of training and readily adaptable to other classification settings, such as training vision models from scratch. Our method is supported by new theoretical results, and requires a low computational budget. We achieve strong empirical results, paving the way for the systematic identification and protection of vulnerable samples before they are memorized.
\end{abstract}
\end{frontmatter}

\section{Introduction} \label{sec:intro}

Large Language Models (LLMs) have revolutionized the way we approach natural language understanding. The availability to the general public of models such as ChatGPT, capable of solving a wide range of tasks without adaptation, has democratized their use. However, a growing body of research have shown that these models memorize a significant proportion of their training data, raising legal and ethical challenges \citep{zhang_understanding_2017, carlini_quantifying_2023, mireshghallah_empirical_2022}. The impact of memorization is ambiguous. On the one hand, it poses a serious threat to privacy and intellectual property because LLMs are often trained on large datasets including sensitive and private information. Practical attacks have been developed to extract this information from training datasets \citep{carlini_extracting_2021, lukas_analyzing_2023, yu_bag_2023, nasr_scalable_2023}, and LLMs have also been shown to plagiarize copyrighted content at inference time \citep{lee_language_2023, henderson_foundation_2024}. On the other hand, memorization can positively impact model's performance, because memorized samples are highly informative. Studies have revealed that outliers are more likely to be memorized, and that these memorized outliers help the model generalize to similar inputs \citep{feldman_does_2020, feldman_what_2020, wang_memorization_2024}.

\begin{figure}[ht!]
    \centering
    \includegraphics[width=\columnwidth]{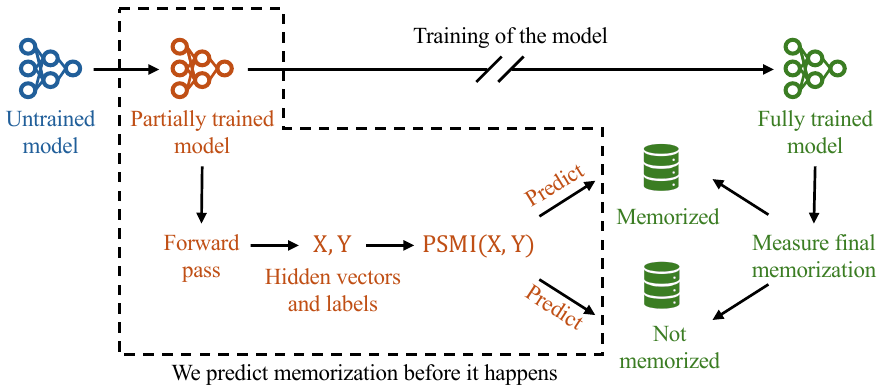}
    \caption{Overview of our method. We interrupt training when the median training loss has decreased by 95\%. We compute a forward pass to retrieve $X$, the hidden representation of the inputs within the partially trained model. We measure the consistency between $X$ and the label $Y$, and use it to predict memorization within the fully trained model.}
    \label{fig:teaser_pipeline}
\end{figure}

Mitigating the negative impacts of memorization while still harnessing its advantages is a challenging task, that requires varying approaches based on the sensitivity of the training data and the purpose of the model. However, practitioners often struggle to evaluate the potential risk of memorization, as empirical defenses often fail to capture the most vulnerable samples from the training set \citep{aerni_evaluations_2024}. To address this limitation, we propose a new method to audit models under development and predict, from the early stages of training, the elements of the training data that the LLM is likely to memorize. This is intended to help practitioners assess the privacy risks of the models they are training, at minimal cost. Existing \textit{a posteriori} measures such as LiRA or counterfactual memorization \citep{carlini_membership_2022, feldman_what_2020} are prohibitively expensive in terms of computational time for most practitioners. In contrast, our method requires no shadow models and can even assess memorization risks before training is complete. The low cost and simplicity of our approach make privacy audits of models under development more widely accessible.

Moreover, our method would enable researchers to design new empirical defenses that optimally allocate privacy budgets to protect the most vulnerable samples. Existing methods such as differential privacy \citep{dwork_calibrating_2006} allocate privacy budgets uniformly. Our method can serve as a proxy for vulnerability to memorization, making it possible to concentrate privacy budgets on the most vulnerable samples before they are memorized, thereby achieving a better privacy-utility trade-off.

Predicting memorization \emph{before it occurs} is an underexplored problem. To address it, we relied on metrics previously used in the literature to characterize memorization in neural networks: Loss, Logit Gap, Mahalanobis distance \citep{mahalanobis_generalized_1936}, and Pointwise Sliced Mutual Information (PSMI) \citep{goldfeld_sliced_2021, wongso_pointwise_2023}. However, these metrics have never been studied for predicting memorization before it occurs. In this paper, we develop and implement a rigorous empirical protocol to evaluate the predictive power of these established metrics for early memorization detection, across a variety of settings.

To the best of our knowledge, \citet{biderman_emergent_2023} provide the only baseline suitable for comparison. However, their approach focuses on generative models, while we study models trained on discriminative tasks. Moreover, they rely on \textit{k-extractability} \citep{carlini_extracting_2021}, which is not applicable to discriminative models. To enable comparison, we adapted their method to a classification setting, despite the significant computational cost arising from the increased complexity of measuring memorization in this context (see Section \ref{sec:eval_protocol}). Once adapted, we used this method as a baseline. While it achieves a low False Positive Rate (FPR), it also suffers from a low True Positive Rate (TPR). Conversely, our approach achieves both low FPR and high TPR while being 50 times faster, paving the way for inspecting vulnerable samples under realistic conditions.

\paragraph{Overview of our approach}

To predict memorization before it occurs, we interrupt training when the median training loss has significantly decreased, typically by 95\% (see Figure \ref{fig:teaser_pipeline}). This drop indicates that the model has learned simple patterns in the hidden representations, enabling it to accurately classify the easiest samples, without relying on memorization. At this stage, we measure the consistency between the labels and the hidden representations of the elements within the partially trained model. If a hidden representation fails to adequately explain its assigned label, it indicates that the sample behaves as a local outlier, within the data distribution's long tail \citep{zhu_capturing_2014}. Such outliers are particularly vulnerable to memorization, because the model will likely fail to learn meaningful representations for them, and will instead resort to memorizing them \citep{feldman_what_2020}. We evaluated four metrics to quantify the consistency between the hidden representations and the labels, with the objective of predicting memorization in the fully trained model. With the exception of Mahalanobis distance, all metrics achieved strong empirical results. 

\begin{figure}[t!]
    \centering
    \includegraphics[width=0.49\columnwidth]{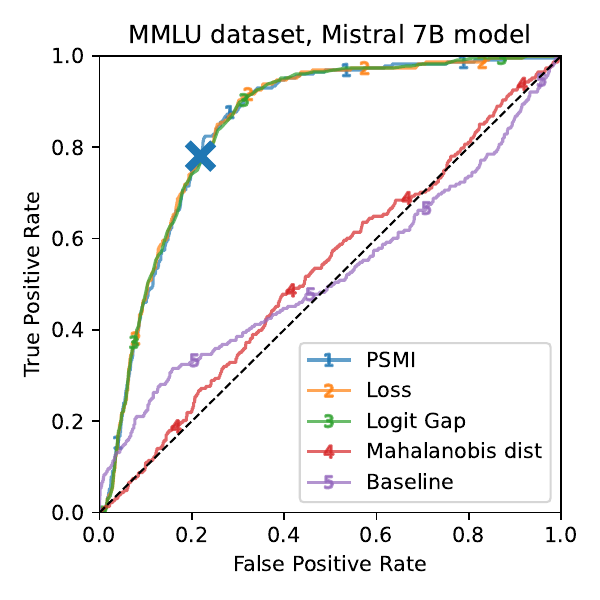}
    \includegraphics[width=0.49\columnwidth]{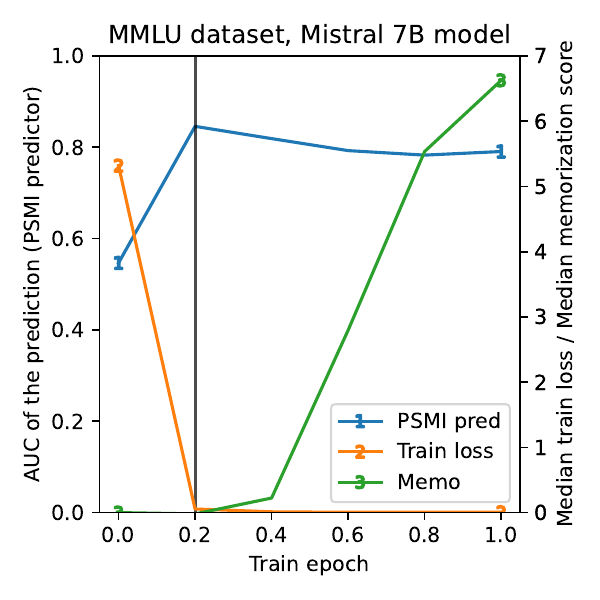}
    \caption{\textbf{Left:} We propose four different metric to predict memorization in the final model: PSMI, Loss, Logit Gap, and Mahalanobis distance. Three of them significantly outperform the baseline, adapted from \citet{biderman_emergent_2023}. The blue cross represents the default PSMI threshold, set to zero (see $\htau$ in Algorithm \ref{alg:predict_memo}). \textbf{Right:} We interrupt training to predict memorization when the median training loss has decreased by 95\% (vertical line). This occurs early in the training pipeline, and predictions at this stage are highly effective (high AUC for "PSMI pred"). Importantly, the samples are not yet memorized at this point (see "Memo" line, which represents the median memorization score of samples memorized by the fully trained model).}
    \label{fig:teaser_results}
\end{figure}

After the prediction, practitioners using our method as a privacy auditing tool should assess whether the risk is acceptable and continue training to completion if so. Indeed, the model used for prediction remains largely under-trained. Importantly, we demonstrate that samples memorized by the fully trained model are not yet memorized at the time of prediction (see Figure~\ref{fig:teaser_results}). This is a key property for researchers aiming to develop new empirical defenses, as it enables to protect vulnerable samples before they are memorized.

\paragraph{Our main contributions can be summarized as follows.}

\begin{itemize}
    \item We propose an algorithm that efficiently predicts, from the early stages of training, whether a sample will be memorized when fine-tuning a LLM for a classification task;
    \item We demonstrate a new theorem to justify our approach;
    \item We provide a PyPI package along with recommended hyperparameter values to facilitate the adoption of our method.
    \item We implement a thorough experimental protocol to compare the effectiveness of several metrics for predicting memorization, using three fine-tuned 7B LLMs and three academic datasets.
    \item We demonstrate the versatility of our approach by applying it as-is to convolutional models trained from scratch on CIFAR-10, yielding strong empirical results.
\end{itemize}
\section{Related work} \label{sec:related_works}

\subsection{Membership Inference Attacks (MIA)}

These attacks were introduced by \citet{shokri_membership_2017}, and aim to determine whether a target sample was part of a target model's training set. Although they are less realistic and practical than extraction attacks \citep{carlini_extracting_2021, lukas_analyzing_2023, nasr_scalable_2023}, membership inference attacks have become the standard approach for measuring the amount of private information a model can leak. Popular attacks such as those of \citet{shokri_membership_2017, carlini_membership_2022, wen_canary_2023} involve training a large number of \textit{shadow models} with different training data. Due to the significant computational resources required, alternative attack methods have been developed that necessitate fewer shadow models or none at all \citep{yeom_privacy_2018, mattern_membership_2023, zarifzadeh_low-cost_2024}.

\subsection{Several definitions of unintended memorization}

For discriminative models, memorization is usually defined as vulnerability to MIA  \citep{mireshghallah_quantifying_2022, carlini_privacy_2022, aerni_evaluations_2024}. Counterfactual memorization can also be applied, requiring the training of multiple models with varying datasets to capture the influence of individual data samples \citep{feldman_what_2020}. On the opposite, to focus on more realistic threats, memorization can be defined as vulnerability to extraction or reconstruction attacks \citep{carlini_secret_2018, carlini_extracting_2021, carlini_quantifying_2023, biderman_emergent_2023, lukas_analyzing_2023, dentan_reconstructing_2024}. These definitions are mostly used with generative models, as such attacks are more complex to implement on discriminative models and often achieve lower performance. However, a large majority of elements extracted consist of common strings frequently repeated in standard datasets \citep{lee_deduplicating_2022, prashanth_recite_2024}. This is why counterfactual memorization was adapted to generative models \citep{zhang_counterfactual_2023, wang_memorization_2024, pappu_measuring_2024, lesci_causal_2024}. Finally, MIA can also be used for generative models  \citep{meeus_copyright_2024}.

\subsection{Explaining and predicting memorization}

Memorization has been commonly associated with overfitting and considered the opposite of generalization. However, this belief was challenged by \citet{zhang_understanding_2017}, who proved that a model can simultaneously perfectly fit random labels and real samples. This phenomenon was studied further in \citep{arpit_closer_2017, chatterjee_learning_2018}, followed by \citet{feldman_does_2020} who explained theoretically how memorization can increase generalization. His idea is that a substantial number of samples in typical datasets belong to the long tail of the distribution \citep{zhu_capturing_2014}, and behave like local outliers, unrepresentative of the overall distribution. As a result, memorizing them enables the model to generalize to similar samples. \citet{feldman_what_2020} and \citet{zhang_counterfactual_2023} confirmed this idea empirically, by observing that memorized samples are relatively difficult for the model. Similarly, it was observed that memorization in self-supervised learning can increase generalization~\citep{wang_memorization_2024}.

A different approach to explain memorization is to analyze the hidden representations learned by the model. For example, \citet{azize_how_2024} linked the privacy leakage of a sample to the Mahalanobis distance \citep{mahalanobis_generalized_1936} between the sample and its data distribution. \citet{leemann_is_2024} evaluated several metrics to predict memorization from a reference model, and concluded that test loss is the best predictor. \citet{wongso_using_2023} computed Sliced Mutual Information \citep{goldfeld_sliced_2021} between the hidden representations and the labels. They theoretically explain why a low SMI indicates memorization, and successfully observed this phenomenon in practice.

These approaches provide \textit{a posteriori} explanations of memorization, because they are either computed from the fully trained model or from a reference model. On the opposite, \citet{biderman_emergent_2023} introduced a new method to predict memorization \textit{before} the end of pre-training. They achieve promising results with high accuracy. However, they obtain low recall scores, indicating that a significant proportion of the samples that are memorized by the final model cannot be detected using their metrics. As they acknowledge, this is an important shortcoming of their method.
\section{Problem setting} \label{sec:problem_setting}

\subsection{Threat model: predicting memorization} \label{sec:threat_model}

We adapt the threat model of \citet{biderman_emergent_2023}. We assume that an engineer is planning to fine-tune a LLM on a private dataset for a classification task, where a small proportion of the dataset contains sensitive information that should not be memorized by the model for privacy concerns. The engineer has full access to the model, its training pipeline and intermediate checkpoints. They do not have the computational budget to train the shadow models required for \textit{a posteriori} measures of memorization such as LiRA \citep{carlini_membership_2022} or counterfactual memorization \citep{feldman_what_2020}. Consequently, they aim to conduct some tests at the beginning of the full training run to approximate \textit{a posteriori} memorization, and determine if the sensitive samples will be memorized by the fully trained model (see Figure \ref{fig:teaser_pipeline}). The engineer wishes to dedicate only a small amount of compute for these tests, to reduce the overhead of confidentiality checks. Moreover, they aim to detect vulnerable samples early to inspect them before they are memorized and decide whether to accept the privacy risk, anonymize or remove the samples, or implement mitigation techniques.

\subsection{Scope: LLMs for classification} \label{sec:paper_scope}

Most studies on memorization in classification settings focus on models of intermediate size trained on datasets such as CIFAR-10 or CIFAR-100 \citep{aerni_evaluations_2024, carlini_privacy_2022, feldman_what_2020}. We have decided to consider more recent scenarios using LLMs fine-tuned for classification tasks. Indeed, generative models are increasingly trained to produce formatted outputs for tasks traditionally handled by discriminative models, such as information extraction \citep{kim_ocr-free_2022, dhouib_docparser_2023}, sentiment analysis \citep{smid_llama-based_2024}, or recommendation \citep{geng_recommendation_2022, cui_m6-rec_2022}. This trend is further amplified in multi-agent systems, which increasingly rely on models fine-tuned to produce short, structured responses as agents profiled for a specific tasks~\citep{wang_survey_2024}. Therefore, we focused on LLMs fine-tuned for classification. However, our method relies on the specific properties of neither LLMs nor fine-tuning. Consequently, it is suitable for any model trained for classification. In Section~\ref{sec:applying_cifar10}, we apply our method as-is to a convolutional network trained from scratch on CIFAR-10, yielding strong results. These results, obtained without any adaptation in a remarkably different setting, demonstrate the applicability of our method to a broad range of classification scenarios.

\section{Methodology} \label{sec:methodology}

\subsection{Preliminary} \label{sec:preliminary}

\paragraph{Hidden representations in Large Language Models}

We consider a decoder-only transformer-based LLM such as Llama 2 \citep{touvron_llama_2023-1} trained on a Multi-Choice Question (MCQ) dataset such as MMLU \citep{hendrycks_measuring_2021}. All tokens of the input are embedded into \textit{hidden representations} in $\RR^{d}$. They are successively transformed at each of the $K$ layers to incorporate information from the context. Finally, the representation of the last token at the last layer is used to predict the label. 

For $k \in \llbracket 1, K \rrbracket$, let $X_k \in \RR^{d}$ be the hidden representation of the last token after the $k$-th layer, and $Y \in \{0, 1, 2, \dots, r\}$ the answer of the MCQ. We can think of $X_k$ and $Y$ as random variables following a joint probability distribution $\mathcal{D}_k$ that can be estimated from the dataset. In the following, we use information-theoretic tools to analyze the interplay between variables $X_k$ and $Y$. Note that $\mathcal{D}_k$ and $X_k$ depend of the training step, but we omit this aspect in our notations to consider a LLM that we freeze to analyze its representations.

\paragraph{(Pointwise) Sliced Mutual Information}

Sliced Mutual Information (SMI) was introduced by \citet{goldfeld_sliced_2021}. Similar to Mutual Information (denoted $\MI$), it measures the statistical dependence between two random variables such as $X_k$ and $Y$. Intuitively, it measures how much the realization of $X_k$ tells us about the realization of $Y$. If they are independent, it is zero ; and if $X_k$ fully determines $Y$, it is maximal. Here, it represents how useful the hidden representations are to predict the labels: we expect SMI to increase with $k$ as the representations become more effective over layers.

\begin{definition}
    Sliced Mutual Information ($\SMI$) is the expectation of Mutual Information (denoted $\MI$) over one-dimensional projections sampled uniformly at random on the unit sphere (denoted $\mathcal{U}(\mathbb{S}^d)$):

\begin{equation}
\begin{gathered}
    \SMI(X_k, Y) 
    = \Exp{\theta \sim \mathcal{U}(\mathbb{S}^d)} \quad 
    \left[
        \MI(\theta^TX_k, Y)
    \right] \\
    = \Exp{\theta \sim \mathcal{U}(\mathbb{S}^d)} \quad
    \left[
        \Exp{(X_k, Y) \sim \mathcal{D}_k} \quad
        \left[
            \log \frac{p(\theta^TX_k, Y)}{p(\theta^TX_k)p(Y)}
        \right]
    \right]
\end{gathered}
\end{equation}
\end{definition}

Pointwise Sliced Mutual Information (PSMI) was introduced as an explainability tool in computer vision \citep{wongso_pointwise_2023}. For every individual realization $(x_k, y)$ of variables $(X_k, Y)$, it represents how surprising it is to observe $x_k$ and $y$ together. For example, a low PSMI means that label $y$ was unexpected with $x_k$, maybe because all similar representations to $x_k$ are associated with another $y' \neq y$ in the dataset.

\begin{definition} \label{def:psmi}
    Pointwise Sliced Mutual Information ($\PSMI$) is defined for every realization $(x_k, y) \in \RR^d \times \llbracket 0; r \rrbracket$ of variables $(X_k, Y)$ as:

\begin{equation} \label{eq:psmi_definition}
    \PSMI(x_k, y) 
    = \Exp{\theta \sim \mathcal{U}(\mathbb{S}^d)} \quad
    \left[
        \log \frac{p(\theta^Tx_k, y)}{p(\theta^Tx_k)p(y)}
    \right]
\end{equation}

\end{definition}

Here, $p$ represents the value of the probability distribution function. It depends on the joint distribution $\mathcal{D}_k$, and can be estimated numerically by approximating $p(\theta^Tx_k \given y)$ by a Gaussian \citep{wongso_pointwise_2023}. The resulting estimator of PSMI is very fast to compute and easy to parallelize. The bottleneck is to compute the hidden representations $x_k$, which requires one forward pass per sample.

\subsection{Defining and measuring memorization} \label{sec:defining_memo}

We adopt LiRA membership inference attack \citep{carlini_membership_2022} as our ground truth memorization, a common choice in classification settings \citep{carlini_privacy_2022, aerni_evaluations_2024} because extractability is rarely used for discriminative models. We empirically verified that LiRA is highly correlated with counterfactual memorization in our setting, confirming its suitability as a ground truth for memorization (see Supplement, Section~\ref{app:comparing_memorization}). LiRA can take very high and low positive values. For convenience, we always use its natural logarithm. Unless otherwise stated, we define memorized samples as those with log-LiRA $\ge 4$, corresponding to strong memorization: the attack estimates that the sample is $e^4 \simeq 54.6$ times more likely to be a member than a non-member of the training set.

\begin{definition} \label{def:memo}
    A sample $x$ is said to be memorized by a model $f_{S^*}$ if:

\begin{equation}
    \log \left[ \text{\normalfont LiRA}(x, f_{S^*}) \right] \ge \heta \coloneqq 4
\end{equation}
\end{definition}

In our experiments, we used 100 shadow models for LiRA, consistent with the findings of \citet{carlini_membership_2022}. In Section~\ref{sec:abl_eta}, we evaluate the impact of choosing threshold $\heta = 4$ on our empirical results.

\subsection{Method and Hyperparameter Choices} \label{sec:our_method}

\begin{algorithm}[h!]
\caption{Predicting Memorization Before It Occurs}\label{alg:predict_memo}
\begin{algorithmic}[1]
\State Interrupt training once the median training loss has decreased by a \textbf{factor} $\hrho$ from its initial value.
\State For each sample $x$, compute a forward pass to extract hidden representations after \textbf{layer} $\hlambda$.
\State For each sample $x$, compute \textbf{metric} $\bm{\mu(x)}$ using hidden vectors.
\State Predict that samples with $\bm{\mu(x) \le \tau}$ will be memorized.
\State Resume training until completion.
\end{algorithmic}
\end{algorithm}

We propose Algorithm~\ref{alg:predict_memo} to predict memorization before it occurs. The goal of $\hmu$ is to identify samples that are hard to classify, as these are more likely to be memorized. We propose four metrics, evaluated in Section~\ref{sec:experiments}: Loss, Logit Gap, Mahalanobis Distance, and PSMI. Since the Loss and Mahalanobis Distance increase for harder-to-classify points, which are more likely to be memorized, we use $\mu(x) = -1 \times \text{Loss}(x)$ (and similarly for Mahalanobis Distance) to maintain consistency with the $\le$ operator in line 4 of Algorithm~\ref{alg:predict_memo}.

\paragraph{Hyperparameters for practitioners}

For practitioners seeking a ready-to-use solution to audit their model with minimal adaptation, we recommend using the values from the first row of Table~\ref{tab:hpar_values}. We interrupt training when the median training loss has decreased by 95\% and measure PSMI at the latest layer. A key advantage of PSMI is that the value $\htau = 0$ has proven robust across a wide range of scenarios and is supported by theoretical justification (see Sections~\ref{sec:psmi_theorem} and \ref{sec:experiments}). This is especially useful for practitioners, who often lack the time or computational budget to optimize $\htau$ and need a reliable, ready-to-use metric to audit privacy risks. To simplify implementation, we provide a PyPI package including an automated estimator of PSMI.\footnote{\safelink{https://pypi.org/project/psmi/}{PSMI PyPI}}

\paragraph{Hyperparameters for researchers}

For researchers seeking to use our method as a proxy for vulnerability to memorization in order to develop new empirical defenses, we recommend using the second row of Table~\ref{tab:hpar_values}. Indeed, Loss can outperform PSMI when $\htau$ is carefully optimized, and researchers typically have the expertise and computational budget to perform such optimization.

\begin{table}[t!]
\centering
\begin{adjustbox}{}
\begin{tabular}{lcccc}
\toprule
\textbf{Profile} & $\hrho$ & $\hlambda$ & $\hmu$ & $\htau$ \\
\midrule
Practitioners & 95\% & latest & $\PSMI$ & $0$ \\
Researchers & 95\% & - & $-1 \times$ Loss & To be optimized \\
\bottomrule
\end{tabular}
\end{adjustbox}
\caption{Hyperparameter recommendations, by user profile}
\label{tab:hpar_values}
\end{table}

\subsection{Theoretical justification} \label{sec:psmi_theorem}

It has been demonstrated that outliers and hard-to-classify points are likely to be memorized \citep{feldman_does_2020, feldman_what_2020}. Here, we show in a simplified setting that the expected value of the PSMI of such points is null, while it is positive for non-outliers. This justifies the use of $\hmu = \PSMI$ with $\htau = 0$ as default parameters for practitioners in Table \ref{tab:hpar_values}. We consider a balanced binary classification with outliers. With probability $1 - \varepsilon$, the point is not an outlier, and the hidden representation $X$ follows a Gaussian distribution (Eq. \ref{eq:thm1_hyp1}). This Gaussian behavior is a classical hypothesis derived from the central limit theorem applied to deep neural networks \citep{matthews_gaussian_2018}. Conversely, with probability $\varepsilon$, the point is an outlier, and $Y$ is sampled at random (Eq. \ref{eq:thm1_hyp2}). We prove that on average PSMI is positive for non-outliers (Eq. \ref{eq:thm1_ccl1}), and zero for outliers (Eq.~\ref{eq:thm1_ccl2}).

\newcounter{savedequationbeforethm1}
\setcounter{savedequationbeforethm1}{\value{equation}}
\newcounter{savedtheorembeforethm1}
\setcounter{savedtheorembeforethm1}{\value{theorem}}
\begin{theorem} \label{thm:psmi_outliers}
Let $(X, Y) \in \RR^d \times \{0, 1\}$ be random variables. We assume that $p(Y = 0) = p(Y = 1) = 0.5$ and that $X$ is a continuous random variable. We also assume that there exist $\mu_0, \mu_1 \in \RR^d$ with $\mu_0 \neq \mu_1$, and $\Sigma_0, \Sigma_1 \in \RR^{d\times d}$, and a Bernoulli variable $\Delta \sim \mathcal{B}(\varepsilon)$ with $\varepsilon \in ]0, 1[$ such that:

\begin{equation} \label{eq:thm1_hyp1}
\left \{ \begin{gathered}
p(X \given Y=0, \Delta=0) \sim \mathcal{N}(\mu_0, \Sigma_0) \\
p(X \given Y=1, \Delta=0) \sim \mathcal{N}(\mu_1, \Sigma_1)
\end{gathered} \right .
\end{equation}

\begin{equation} \label{eq:thm1_hyp2}
\forall x \in \RR^d, \quad 
\left \{ \begin{gathered}
p(Y = 0 \given \Delta=1, X=x) = 0.5 \\
p(Y = 1 \given \Delta=1, X=x) = 0.5
\end{gathered} \right .
\end{equation}

Given this, we then have:
    
\begin{equation} \label{eq:thm1_ccl1}
    \Exp{X, Y} \left[ \PSMI(X, Y) \given \Delta = 0 \right] > 0
\end{equation}
\begin{equation} \label{eq:thm1_ccl2}
    \Exp{X, Y} \left[ \PSMI(X, Y) \given \Delta = 1 \right] = 0
\end{equation}

\end{theorem}

\begin{proofsketch}
We first prove Eq.~\ref{eq:thm1_ccl2} by writing the integral formulation  of PSMI from Eq.~\ref{eq:psmi_definition}. Given $\Delta = 1$, we have $p(\theta^Tx_k, y) = p(\theta^Tx_k)p(y)$ due to Eq.~\ref{eq:thm1_hyp2}, which simplifies the fraction to 1 and leads to an expected value of zero. Then, using Eq.~\ref{eq:thm1_ccl2}, we prove that $\EE[\PSMI(X, Y) \given \Delta = 0] > \SMI(X, Y)$. As a result, it is sufficient to demonstrate Eq.~\ref{eq:thm1_ccl1} with $\SMI(X, Y)$ instead of $\EE[\PSMI(X, Y) \given \Delta = 0]$. To do this, we apply Theorem 1 from \citep{wongso_using_2023}. For that purpose, we propose a method to find $(R_0, R_1, m_g, \nu) \in \RR_{+, *}^4$ such that $(X, Y)$ is $(R_1, R_2, m_g, \nu)$-SSM separated w.r.t. Definition 3 in \citep{wongso_using_2023}. See full proof in Supplement, Section~\ref{app:proof_theorem_1}.
\end{proofsketch}

\section{Experimental results} \label{sec:experiments}

\begin{figure*}[t!]
    \centering
    \includegraphics[width=0.32\textwidth]{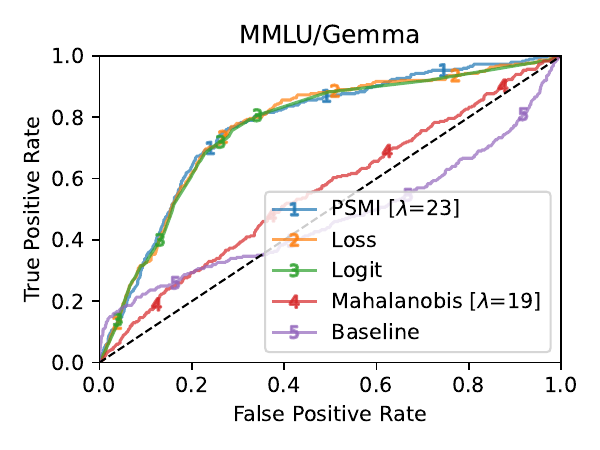}
    \includegraphics[width=0.32\textwidth]{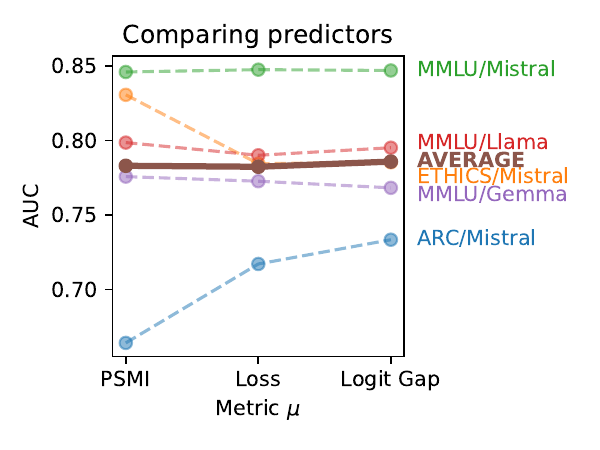}
    \includegraphics[width=0.32\textwidth]{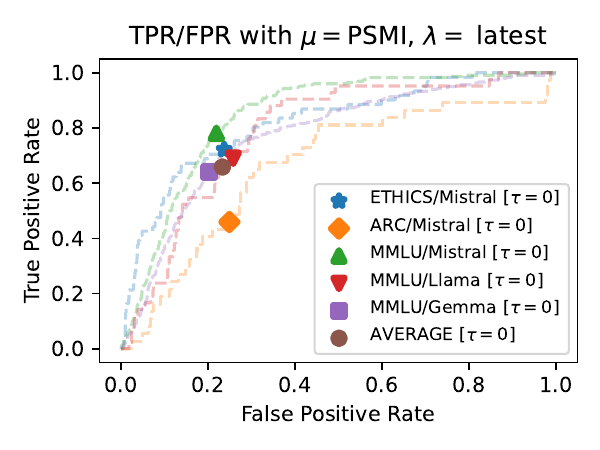}
    \caption{\textbf{Left:} TPR/FPR trade-off for the four metrics $\hmu$ and the baseline. The baseline and Mahalanobis Distance perform near random, while the others are accurate predictors. We use the layer $\hlambda$ that maximizes AUC for each metric. \textbf{Center:} AUC of the best-performing metrics $\hmu$ across different datasets and models. \textbf{Right:} Algorithm~\ref{alg:predict_memo} with predictor $\hmu = \PSMI$. Markers indicate performance at $\htau = 0$. Dashed lines show the TPR/FPR trade-off when $\htau$ varies. }
    \label{fig:pred_perfs}
\end{figure*}
\subsection{Evaluation protocol} \label{sec:eval_protocol}

\paragraph{Datasets, Models and Training Pipeline}

We used three pretrained models with similar architectures: Mistral 7B v1 \citep{jiang_mistral_2023}, Llama 7B v2 \citep{touvron_llama_2023-1}, and Gemma 7B \citep{gemma_team_gemma_2024}. We used three popular academic benchmarks: MMLU \citep{hendrycks_measuring_2021}, ETHICS \citep{hendrycks_aligning_2021} and ARC \citep{boratko_systematic_2018}. We fine-tuned these models using LoRA \citep{hu_lora_2022} with 0.29\% of trainable parameters and question-answering templates asking the model for the label. Models are trained using Next Token Prediction task, computing the loss only for the token corresponding to the label. To ensure fair \mbox{comparisons} and prevent over-training, we always terminate training after one epoch. We assess five dataset/model combinations: ARC/Mistral 7B, ETHICS/Mistral 7B, MMLU/Mistral 7B, MMLU/Llama 2 7B, and MMLU/Gemma 7B. These combinations allows us to evaluate several datasets with a fixed model (Mistral), and several models with a fixed dataset (MMLU), while keeping computational cost low. The total computational cost amounts to 10,961 hours on Nvidia A100 80G GPUs and 5,787 single-core hours on Intel Xeon 6248 CPUs, corresponding to 0.57 tCO$_2$eq for this specific cluster.\footnote{\safelink{https://www.edari.fr/info-bilan-ges}{CO$_2$ emission sources}}

\paragraph{Evaluation metrics: FPR at high TPR}

We evaluate the effectiveness of our method by measuring ground-truth memorization at the end of training (Definition \ref{def:memo}). We compute the True Positive Rate (TPR) / False Positive Rate (FPR) curve by varying prediction threshold $\htau$. The TPR represents the proportion of memorized samples in the final model that are correctly detected based on the partially trained one, and the FPR represents the proportion of non-memorized samples that are wrongly detected. We prefer TPR / FPR to precision~/~recall because it is independent of the prevalence of memorized samples. However, a high TPR is more important than a low FPR. Indeed, false positives lead practitioners to be overly cautious. This is unprofitable, but it does not threaten privacy. Conversely, false negatives lead practitioners to underestimate memorization, which entails a privacy risk. Consequently, we will focus on regions that achieve a high TPR, typically greater than 75\%. The Area Under the Curve (AUC) provides a single numerical value for comparing metrics $\hmu$, although it presents a simplistic view of the TPR / FPR trade-off.

\paragraph{Baseline: an adaptation of the approach of \citet{biderman_emergent_2023}} We compare our results to their method, because it is the only comparable approach we are aware of. We replaced $k$-extractability with LiRA because our models are trained for classification (see Section~\ref{sec:related_works}). Despite this adaptation, it behaves similarly to the results observed by \citet{biderman_emergent_2023}, with a low FPR at the cost of an insufficiently high TPR. The computational cost of this baseline is significantly higher than that of our method, due to the shadow models required for LiRA. As a result, this baseline is not suitable for practitioners within our threat model. Nevertheless, we used it to compare our results because it is the only comparable approach.

\begin{figure*}[t!]
    \centering
    \includegraphics[width=0.32\textwidth]{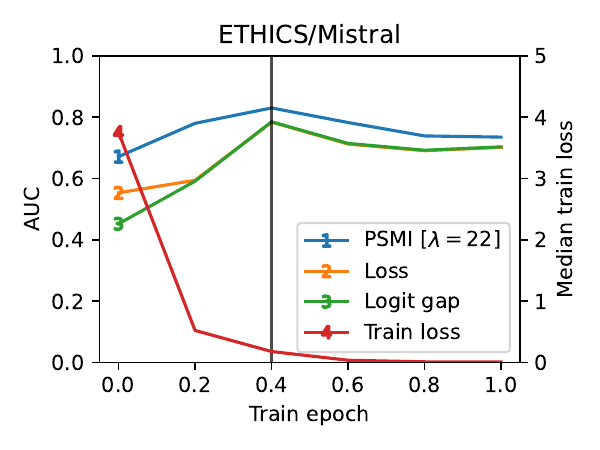}
    \includegraphics[width=0.32\textwidth]{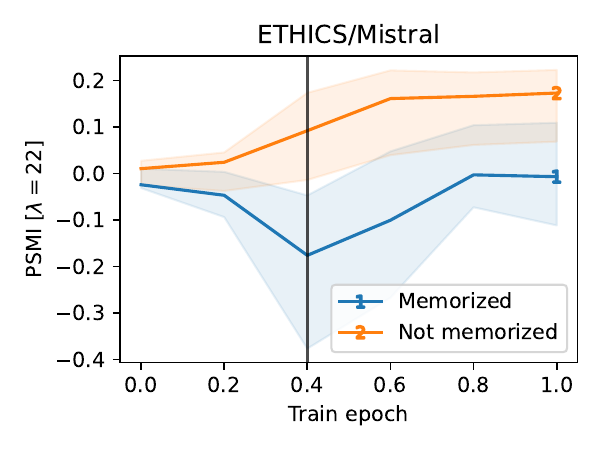}
    \includegraphics[width=0.32\textwidth]{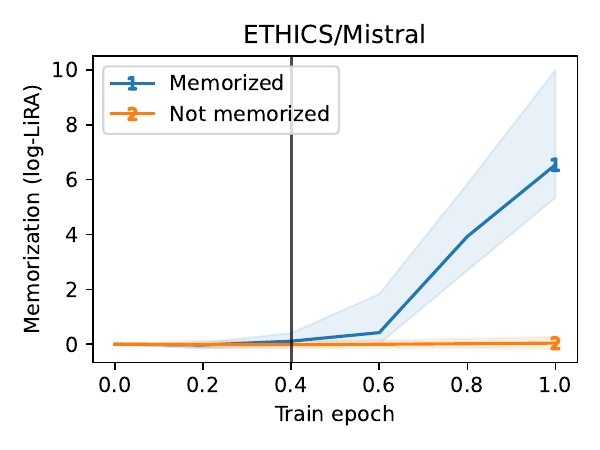}
    \caption{Memorized samples can be detected from epoch 0.4, though they are not yet memorized. \textbf{Left:} Lines 1, 2, 3: AUC of PSMI, Loss and Logit Gap. Line 4: median training loss. The vertical line marks the prediction checkpoint with $\hrho=95\%$. \textbf{Center:} Solid lines show the median PSMI, with the 25\%–75\% quantiles shaded. Samples are grouped into memorized and non-memorized based on the fully trained model. \textbf{Right:} Similar representation of memorization.}
    \label{fig:ablation_rho}
\end{figure*}
\begin{figure*}[t!]
    \centering
    \includegraphics[height=120pt]{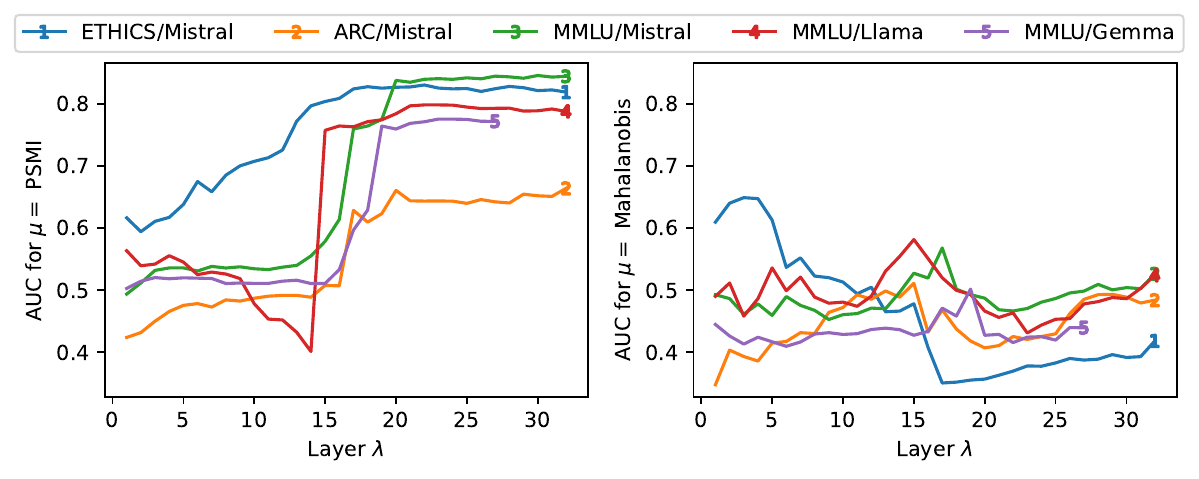}
    \includegraphics[height=120pt]{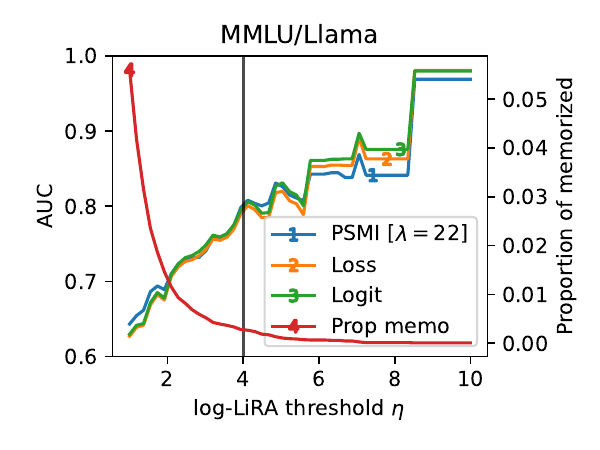}
    \caption{\textbf{Left and Center:} Impact of the layer $\hlambda$ on the AUC of PSMI (Left) or Mahalanobis (Center). \textbf{Right:} Impact of the threshold $\heta$ used in Definition~\ref{def:memo}. The default value is $\heta = 4$ (vertical line). \mbox{Lines 1, 2, 3:} AUC of PSMI, Loss, and Logit Gap. Line 4: proportion of memorized samples as a function of $\heta$.}
    \label{fig:ablation_lambda_and_eta}
\end{figure*}
\subsection{Main results and computational gains} \label{sec:main_results}

Our experiments demonstrate that memorization can be accurately predicted from the early stages of training, and that our method significantly outperforms the baseline. By default, our figures present results with $\heta = 4$ (Definition \ref{def:memo}), $\hrho = 95\%$ and $\hlambda$ optimized to maximize the AUC. We present the TPR/FPR curve or its AUC for each predictor $\hmu$, computed by varying the threshold $\htau$. The impact of each of these parameters is discussed in the following sections. 

\paragraph{Computational Gains: 50x Speed-Up Over Baseline}

The bottleneck of our method is computing a forward pass for every sample, which costs as much as $1/3$ of an epoch~\citep{hobbhahn_whats_2021}. In our settings, prediction is done after a fraction $r = 0.2$ to $0.4$ epoch, resulting in a total cost of $r + 1/3 \simeq 2/3$ epoch. On the opposite, both ground truth memorization and the baseline require shadow models. With 100 shadow models, the ground truth costs 100 epochs, and the baseline costs $100 \times r$ epochs. Consequently, our approach is 150 times faster than the ground truth, and 50 times faster than the baseline. This significant reduction in computational cost is crucial, as it makes accessible an estimate of memorization that was previously unattainable with the typical budget of a practitioner within our threat model.

\subsection{Impact of the choice of predictor $\hmu$} \label{sec:abl_mu}

Figure \ref{fig:pred_perfs} (Left) compares several predictors $\hmu$ for detecting memorization on MMLU/Gemma. Three metrics significantly outperform the baseline: PSMI, Loss, and Logit Gap. They achieve high AUC values: 77.6\%, 77.2\%, and 76.8\%, respectively. In Figure~\ref{fig:pred_perfs} (Center), we compare the AUC of the three strong predictors across five dataset/model combinations. On average, they achieve very similar AUC. In Table \ref{tab:hpar_values}, we recommend researchers to use the loss, as it is the simplest to implement, but any of the three metrics is suitable within our threat model and yields strong results. See Figure \ref{fig:appendix_mu} in the Supplement for TPR/FPR curves across other datasets/models.

\subsection{Impact of the prediction threshold $\htau$} \label{sec:abl_tau}

Parameter $\htau$ allows us to move along the TPR/FPR curve to accommodate the specific needs of each situation. We expect researchers to optimize $\htau$ accordingly. However, to enable broader adoption of privacy audits by practitioners, we need to provide a ready-to-use value for $\htau$. This is where the metric $\hmu = \PSMI$ offers a key advantage over the others: as suggested by Theorem \ref{thm:psmi_outliers}, setting $\htau = 0$ consistently yields strong results, with an average FPR of 23.3\% and TPR of 65.8\% (Figure \ref{fig:pred_perfs}, Right). These values are well-balanced, with a TPR high enough to capture most vulnerable samples while maintaining a low FPR. We therefore recommend using $\hmu = \PSMI$ and $\htau = 0$ for practitioners seeking a ready-to-use solution to audit their models (see first row in Table \ref{tab:hpar_values}).

\subsection{Impact of the timing of measure $\hrho$} \label{sec:abl_rho}

To select an appropriate value for $\hrho$, we save the models every 0.2 epochs and evaluate how effectively memorization can be predicted at each checkpoint. Figure \ref{fig:ablation_rho} presents the results for ETHICS/Mistral; results for other datasets/models are shown in Figure \ref{fig:appendix_rho} in the Supplement. We observed that $\hrho = 95\%$ yields robust results across our five dataset/model combinations. It also proved effective in a completely different setting (see Section \ref{sec:applying_cifar10}), validating its use as the recommended value in Table \ref{tab:hpar_values}. As shown in Figure \ref{fig:ablation_rho} (Left), the prediction marked by the vertical line does not occur too early. Before that checkpoint, the patterns learned by the model are not yet relevant enough to determine whether a sample will be hard to learn and therefore memorized. This is evident from the clear gap between the PSMI of memorized and non-memorized samples at the prediction checkpoint in Figure \ref{fig:ablation_rho} (Center). Moreover, the prediction does not occur too late, as shown in Figure \ref{fig:ablation_rho} (Right). Memorized samples have not yet been memorized, so we can still implement mitigation techniques, which is an essential requirement of our threat model.

\subsection{Impact of the layer $\hlambda$} \label{sec:abl_lambda}

In Figure \ref{fig:ablation_lambda_and_eta} (Left and Center), we observe that the difference between the AUC of PSMI using the last layer and the AUC using the best layer is minor, which supports recommending the last layer as a robust choice in Table \ref{tab:hpar_values}. We also see that the AUC with Mahalanobis distance is consistently lower than that of PSMI, which corroborates our discussion in Section \ref{sec:abl_mu}. This performance gap can be attributed to the fact that PSMI incorporates more information: while Mahalanobis distance measures the similarity of a sample to a single data distribution, PSMI accounts for both $p(\theta^Tx_k \given y)$ and the distribution over other labels via $p(\theta^Tx_k)$ (see Definition~\ref{def:psmi}).

For PSMI, we also observe that the importance of $\hlambda$ varies across datasets. Curves 1, 2, and 3 show the same model for different datasets. For MMLU and ARC, the curve rises sharply around layers 15–20 and then stabilizes, while for ETHICS, it increases more smoothly. Our interpretation is that MMLU and ARC are complex reasoning tasks, requiring more intricate interactions between token representations, which concentrate relevant layers toward the end of the network. In contrast, ETHICS is a simpler binary classification task, where samples are easier to separate with fewer token interactions, allowing memorization to be detected from earlier layers. Conversely, curves 3, 4, and 5 show that for a fixed dataset, the choice of model has little impact on the shape of the curve.

\begin{figure*}[t!]
    \centering
    \includegraphics[width=0.32\textwidth]{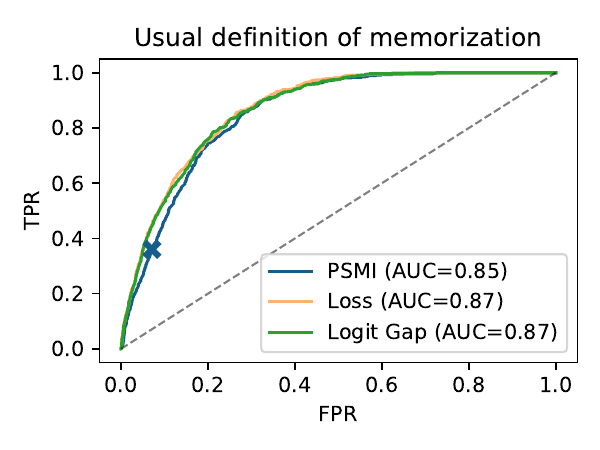}
    \includegraphics[width=0.32\textwidth]{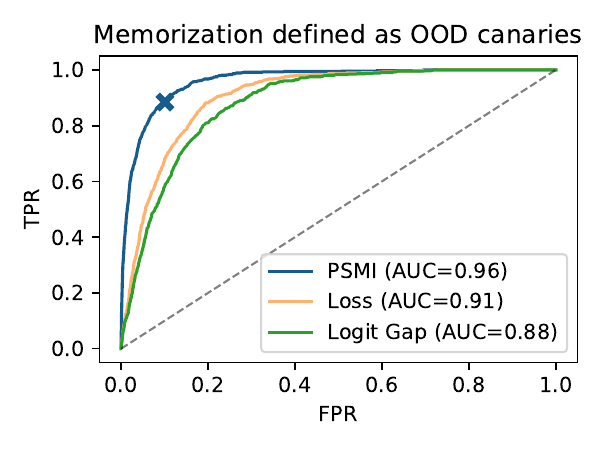}
    \includegraphics[width=0.32\textwidth]{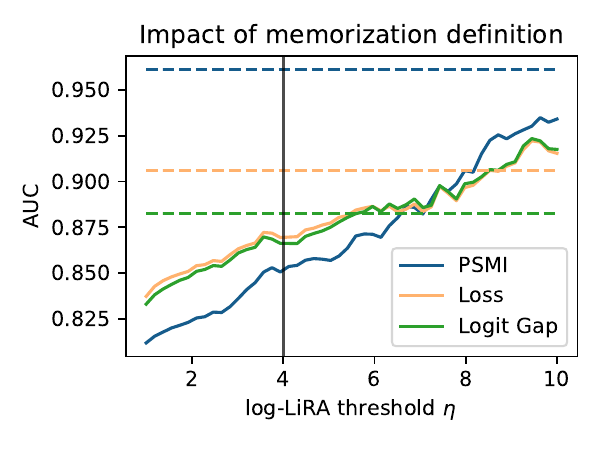}
    \caption{Applying our method as-is on a WRN16-4 \citep{zagoruyko_wide_2016} trained for 300 epochs on CIFAR-10. Using $\hrho = 95\%$, the prediction is made after only 4 epochs. \textbf{Left:} TPR/FPR of PSMI ($\hlambda = $ latests), Loss and Logit Gap. The blue cross marks the performance of the recommended parameters for practitioners (see Table \ref{tab:hpar_values}). \textbf{Center:} Same experiment, but memorized samples are defined as OOD canaries to mimic the most vulnerable samples (see \citet{aerni_evaluations_2024}) \textbf{Right:} The solid line represents the impact of $\heta$ on the AUC. The dashed line represents the AUC when memorized samples are defined as the OOD canaries.}
    \label{fig:ablation_cifar}
\end{figure*}
\subsection{Impact of the log-LiRA threshold $\heta$} \label{sec:abl_eta}

Figure \ref{fig:ablation_lambda_and_eta} (Right) evaluates the impact of $\heta$ on our results for MMLU/Llama; results for other datasets/models are shown in Figures \ref{fig:appendix_eta} and \ref{fig:appendix_memo_histogram} in the Supplement. For MMLU/Llama, 2.8\% of samples are considered memorized with $\heta = 4$. By definition, this proportion decreases as $\heta$ increases. Conversely, we observe that the AUC of our three effective predictors increases with $\heta$. Our interpretation is that the samples most clearly detected as memorized by LiRA were harder for the model to learn, making them easier for our method to identify. This is an important result, as it shows that our method is particularly effective at detecting the most vulnerable samples, which is a desirable property for a privacy auditing tool.

\subsection{Applicability to other classification scenarios} \label{sec:applying_cifar10}

Our method relies on the specific properties of neither LLMs nor fine-tuning. Consequently, it is applicable to any model trained for classification tasks. To validate this hypothesis, we applied our method as-is to a setting entirely different from the fine-tuned LLMs studied so far: a Wide Residual Network (WRN16-4) \citep{zagoruyko_wide_2016} trained from scratch on CIFAR-10, with only 2.7M parameters, without LoRA, and for 300 epochs.

We applied our evaluation protocol to the framework of \citet{aerni_evaluations_2024}. In Figure~\ref{fig:ablation_cifar} (Left), we measure the TPR/FPR trade-off using our three effective predictors and the default memorization definition (Definition~\ref{def:memo}). With hyperparameters $\hrho = 95\%$ and $\hlambda =$ latest, prediction is made at epoch 4 out of 300, yielding excellent AUC for every predictor, even surpassing those obtained with fine-tuned LLMs. Moreover, \citet{aerni_evaluations_2024} introduced out-of-distribution (OOD) \textit{canaries} into the training set and demonstrated that they correctly mimic the samples most vulnerable to membership inference attacks. We repeated our experiments using these canaries instead of Definition~\ref{def:memo}, obtaining particularly strong results (Figure~\ref{fig:ablation_cifar}, Center). These canaries can be viewed as the limit of Definition~\ref{def:memo} as $\heta \rightarrow +\infty$ to focus the most vulnerable samples. This supports our discussion in Section~\ref{sec:abl_eta}. As shown in Figure~\ref{fig:ablation_cifar} (Right), our method performs better at detecting the most vulnerable samples, as evidenced by the increasing AUC and the strong performance on the canaries.

The strong results we obtained on both real CIFAR-10 samples and canaries indicate that our method remains effective in classification settings for which it was neither designed nor optimized.

\section{Discussion}

\subsection{Limitations and future work} \label{sec:limitations}

Our method is specifically designed for classification tasks, and most of our experiments focus on LLMs fine-tuned for multiple-choice questions (see Section \ref{sec:paper_scope}). It would be interesting to investigate whether it can be adapted to LLMs trained on generative tasks. This setting is widely used and presents significant privacy risks, making our method particularly valuable if successfully extended.

Finally, our results could contribute to the development of new empirical defenses. Several methods have already been proposed to mitigate unintended memorization in practice~\citep{chen_relaxloss_2022, tang_mitigating_2022, chen_overconfidence_2024}. Our algorithm could be employed to improve these techniques by optimally allocating the privacy budgets to protect the most vulnerable samples, resulting in a better privacy-utility trade-off.

\subsection{Open science} \label{sec:open_science}

In addition to the hyperparameter discussed in this paper, we provide the following repository containing the Python, Bash and Slurm scripts needed to reproduce our experiments. Moreover, to facilitate the implementation of our method for auditing models under development, we provide a PyPI package with an automated PSMI estimator applicable in a wide range of scenarios.

\begin{center}
\safelink{https://github.com/orailix/predict_llm_memorization}{Project GitHub} \\
\safelink{https://pypi.org/project/psmi/}{PSMI PyPI}
\end{center}

\subsection{Ethics considerations} \label{sec:ethics_considerations}

This paper discusses vulnerability to privacy attacks against language models in practical settings. However, our work cannot benefit adversaries with harmful intent, for several reasons. First, our approach requires access to the checkpoint of a partially trained model, which adversaries cannot do. Second, although our work improves our understanding of unintended memorization, it will benefit privacy researchers more than adversaries. Our approach helps identify vulnerable samples but does not help attack them. Moreover, it can help practitioners better audit models under development and understand the privacy risks they entail. Finally, our results can be used to develop new empirical defenses with better privacy/utility trade-offs.
\section{Conclusion} \label{sec:conclusion}

In this work, we demonstrate that it is possible to predict which samples will be memorized by a language model fine-tuned in a classification setting. Our approach significantly outperforms the only existing baseline while being 50 times computationally more efficient. It can be utilized from the early stages of training to protect vulnerable samples before they are memorized by the model.

We provide a theoretical justification for our approach and validate its effectiveness on three different language model architectures fine-tuned on three different classification datasets. Moreover, we demonstrate that our method is easily applicable to other classification scenarios by successfully applying it, without modification, to a vision model trained from scratch. We provide hyperparameters and package implementations to facilitate the adoption of our method in a wide range of classification settings.

This method lays the groundwork for developing effective tools to evaluate models during training, assess their privacy risks, and efficiently prevent unintended memorization.

\begin{ack}
This work received financial support from Crédit Agricole SA through the research chair “Trustworthy and responsible AI” with École Polytechnique. This work was performed using HPC resources from GENCI-IDRIS 2023-AD011014843. We thank Arnaud Grivet Sébert and Mohamed Dhouib for discussions on this paper.
\end{ack}

\bibliography{bibliography/main}

\appendix
\newpage
\section{Comparing definitions of memorization} \label{app:comparing_memorization}

As mentioned in Section \ref{sec:defining_memo}, we used vulnerability to LiRA as our definition of ground truth memorization. To validate this choice, we compare LiRA to another definition of memorization which is popular for classification models: counterfactual memorization \citep{feldman_what_2020}.

In Section \ref{sec:local_lira}, we introduce the original version of LiRA. In Section \ref{app:global_lira}, we introduce a \textit{global} version of LiRA, which is comparable to counterfactual memorization. In Section \ref{app:actual_comparison_lira}, we present the results of our experiments.

\subsection{LiRA: local version} \label{sec:local_lira}

This section presents the original formulation of LiRA attack as introduced by \cite{carlini_membership_2022}. We call this version \textit{local} because it targets a fixed model and its training set.

Let $\rmX = \{(x_i, y_i)\}_{i \in \llbracket 1, N \rrbracket}$ be a training set of $N$ labelled elements. We focus on multi-choice question (MCQ) academic benchmarks such as MMLU \citep{hendrycks_measuring_2021}. Let $S$ be a random variable representing a subset of  elements in $\llbracket 1, N \rrbracket$. Let $\rmX_S = \{(x_i, y_i) \: | \: i \in S\}$ be the corresponding subset of training elements, and $f_S \sim \mathcal{T}(X_S)$ be a model trained on this subset with the randomized training procedure $\mathcal{T}$. Then, let $\mathcal{L}(x, f_S)$ be the logit gap of the evaluation of $x$ with model $f_S$, i.e. the difference between the highest and second-highest logit.

Let fix a target subset $S^*$, a target model $f_{S^*} \sim \mathcal{T}(\rmX_{S^*})$ trained on these elements, and a target element $x \in \rmX$. As every membership inference attack, LiRA aims to determine whether $x$ was in $\rmX_{S^*}$. First, we train a great number of \textit{shadow models} $f_S$ on random subsets of $\rmX$, and evaluates the logit gap $\mathcal{L}(x, f_S)$ for theses shadow models. Then, we gather $\mathcal{L}^\tin = \{\mathcal{L}(x, f_S) \: | \: x \in S\}$, the logit gaps of models that were trained on a subset containing $x$ ; and $\mathcal{L}^\tout$ for models that were trained on a subset that does not contain $x$. We model the two sets $\mathcal{L}^\tin$ and $\mathcal{L}^\tout$ as Gaussian distributions, and compute the probabilities $p^\tin$ and $p^\tout$ of the target logit gap $\mathcal{L}(x, f_{S^*})$ under these distributions. Finally, we define $\LIRA(x, f_{S^*}) = p^\tin / p^\tout$. It takes very high and low values positive values, which is why we use its natural logarithm in our paper. A value greater that $0$ indicates that the sample is memorized, because $p^\tin > p^\tout$. See the default definition adopted in our article in Section \ref{sec:defining_memo}.

\subsection{LiRA: global version} \label{app:global_lira}

To compare LiRA with counterfactual memorization, we must introduce a \textit{global} version of LiRA. It does not target a fixed model. On the opposite, it attacks multiple models trained on a random splits of the same datasets, and measures the attack success rate of LiRA against each samples. This \textit{global} version is used by \citet{carlini_privacy_2022} for example.

We use the same notations as in Section \ref{sec:local_lira}. The Attack Success Rate (ASR) indicates whether a given element $x \in \rmX$ is likely to be memorized by any model trained on a subset $\rmX_S$ with training procedure $\mathcal{T}$. Let $\mathcal{D}$ be the distribution of $S$ corresponding to choosing a random subset of $\lfloor N/2 \rfloor$ elements in $\llbracket 1, N \rrbracket$, meaning that every element is selected with probability 50\%.

\begin{definition}
Let $x \in \rmX$ be a target element in the training set. We define:

\begin{equation}
\ASR(x) = \underset{S \sim \mathcal{D},\; f_S \sim \mathcal{T}(\rmX_S)}{\mathbb{P}} \;\left[ \mathds{1}[p^\tin > p^\tout] = \mathds{1}[x \in \rmX_S] \right]
\end{equation}
\end{definition}

The global LiRA attack represents the likelihood that a sample in a dataset gets memorized by any model trained with a given procedure. As a result, this score is not consistent with our threat model. Indeed, in our threat model we want to audit a \textit{fixed} model, because this is what practitioners do. This is why we did not use the global version in the main body of this paper.

\subsection{Comparing LiRA and counterfactual memorization} \label{app:actual_comparison_lira}

\begin{figure*}[t!]
    \centering
    \includegraphics[height=72pt]{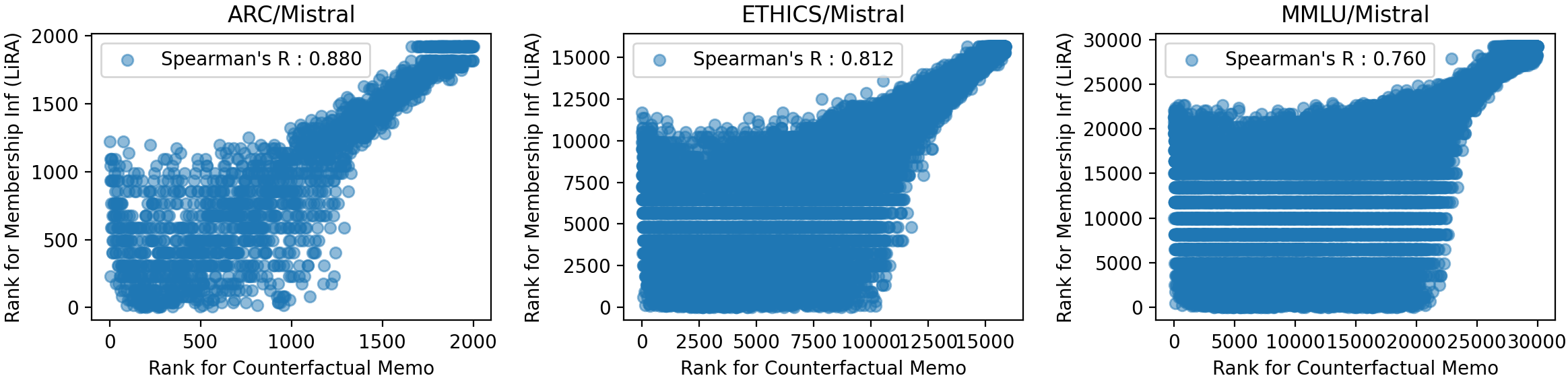}
    \includegraphics[height=72pt]{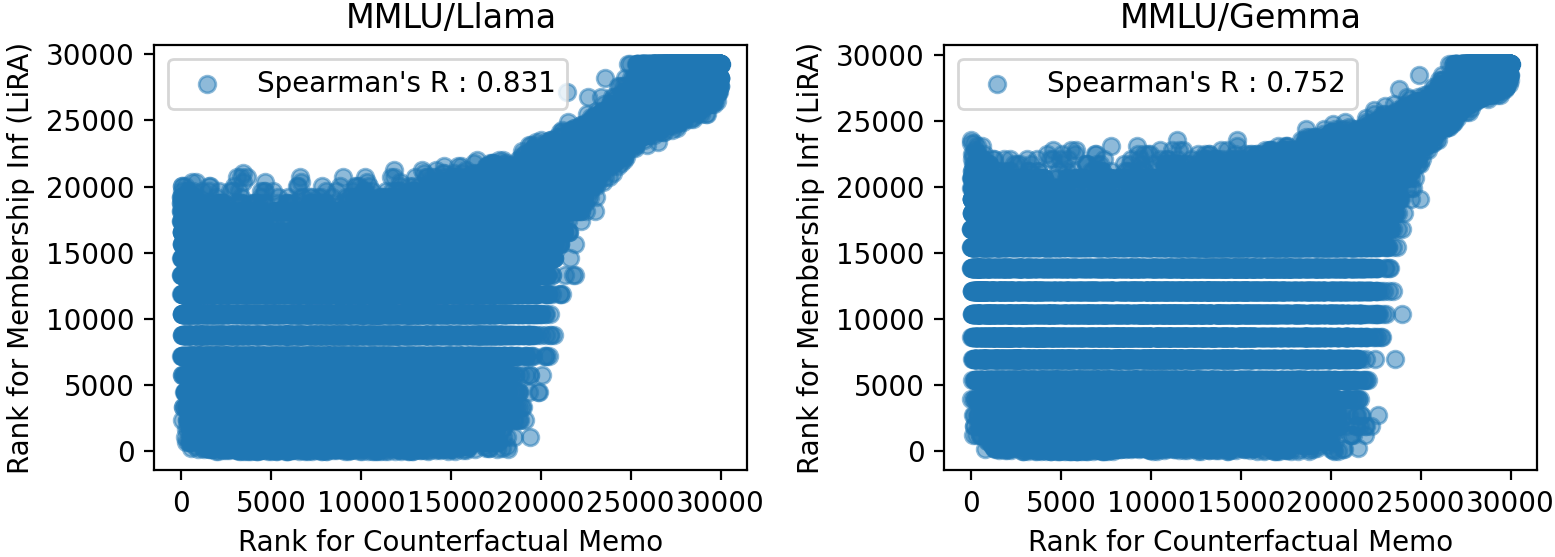}
    \caption{Comparing two definitions of memorization: counterfactual memorization \citep{feldman_what_2020, zhang_counterfactual_2023} and LiRA~\citep{carlini_membership_2022}. We measure Spearmans'R coefficient to evaluate the consistency them. These experiments are conducted on models trained for 10 epochs.}
    \label{fig:appendix_compare_memo}
\end{figure*}

Like the global version of LiRA, counterfactual memorization \citep{feldman_what_2020, zhang_counterfactual_2023} does not measure memorization within a fixed target model, making it inapplicable to our threat model. We evaluate the consistency between counterfactual memorization and the global version of LiRA. We use Equation 2 in \citep{zhang_counterfactual_2023} to compute counterfactual memorization, using the logit gap as the performance metric~$M$.

Our results are presented in Figure \ref{fig:appendix_compare_memo}. We use Spearman's~R score to quantify the consistency between the two definitions, as we are interested in the \textit{order} of samples with respect to each definition. We observe that Spearman's R between the two definitions is high in every setting: it is always greater than 0.75 and increases to 0.88 for ARC/Mistral. This demonstrates that LiRA and counterfactual memorization are consistent with each other. In addition, we can separate the samples into two groups: a first, weakly memorized group, for which there is greater variability between the two definitions (bottom left of the graphs), and a strongly memorized group, for which the two definitions are much more consistent with each other (top right of the graphs). The second group is the most important in our setting, as we are interested in predicting samples that are the most vulnerable to memorization.

The coherence of these two definitions, especially for highly memorized samples, confirms the relevance of choosing LiRA as the ground truth memorization for our experiments in Section \ref{sec:defining_memo}.

\section{Proof of theorem 1 and discussion} \label{app:proof_theorem_1}

In this section we prove Theorem \ref{thm:psmi_outliers} and discuss it. We recall the theorem:

\newcounter{savedequation}
\setcounter{savedequation}{\value{equation}}
\newcounter{savedtheorem}
\setcounter{savedtheorem}{\value{theorem}}
\setcounter{equation}{\value{savedequationbeforethm1}}
\setcounter{theorem}{\value{savedtheorembeforethm1}}
\begin{theorem} \label{thm:psmi_outliers_proof}
Let $(X, Y) \in \RR^d \times \{0, 1\}$ be random variables. We assume that $p(Y = 0) = p(Y = 1) = 0.5$ and that $X$ is a continuous random variable. We also assume that there exist $\mu_0, \mu_1 \in \RR^d$ with $\mu_0 \neq \mu_1$, and $\Sigma_0, \Sigma_1 \in \RR^{d\times d}$, and a Bernoulli variable $\Delta \sim \mathcal{B}(\varepsilon)$ with $\varepsilon \in ]0, 1[$ such that:

\begin{equation} \label{eq:thm1p_hyp1}
\left \{ \begin{gathered}
p(X \given Y=0, \Delta=0) \sim \mathcal{N}(\mu_0, \Sigma_0) \\
p(X \given Y=1, \Delta=0) \sim \mathcal{N}(\mu_1, \Sigma_1)
\end{gathered} \right .
\end{equation}

\begin{equation} \label{eq:thm1p_hyp2}
\forall x \in \RR^d, \quad 
\left \{ \begin{gathered}
p(Y = 0 \given \Delta=1, X=x) = 0.5 \\
p(Y = 1 \given \Delta=1, X=x) = 0.5
\end{gathered} \right .
\end{equation}

Given this, we then have:
    
\begin{equation} \label{eq:thm1p_ccl1}
    \Exp{X, Y} \left[ \PSMI(X, Y) \given \Delta = 0 \right] > 0
\end{equation}
\begin{equation} \label{eq:thm1p_ccl2}
    \Exp{X, Y} \left[ \PSMI(X, Y) \given \Delta = 1 \right] = 0
\end{equation}

\end{theorem}
\setcounter{equation}{\value{savedequation}}
\setcounter{theorem}{\value{savedtheorem}}

\paragraph{Proof of Equation \ref{eq:thm1p_ccl2}} Let $x, y \in \RR^d \times \{0, 1\}$. We use the hypothesis we made in Equation \ref{eq:thm1p_hyp2}:

\begin{align}
    &p(X=x, Y=y \given \Delta = 1) \\
    &= p(Y=y \given \Delta = 1, X=x) \times p(X=x \given \Delta=1) \\
    &= 0.5 \times p(X=x \given \Delta=1) \\
    &= p(Y=y \given \Delta=1) \times p(X=x \given \Delta=1)
\end{align}

Consequently, given $\Delta = 1$, $X$ and $Y$ are independent. We conclude that the expected value of PSMI is zero, which proves Equation~\ref{eq:thm1p_ccl2}.

\begin{align}
    &\Exp{X, Y}[\PSMI(X, Y) \given \Delta = 1] \\
    &=\  \int\limits_{X, Y} \int\limits_{\theta \sim \mathcal{U}(\mathbb{S}^d)} 
        \log \frac{p(\theta^Tx, y)}{p(\theta^Tx)p(y)}
     \diff p(X, Y \given \Delta=1) \diff p(\theta) \\
     &=\ \int\limits_{X, Y} \int\limits_{\theta \sim \mathcal{U}(\mathbb{S}^d)} 
        \log \frac{p(\theta^Tx)p(y)}{p(\theta^Tx)p(y)}
     \diff p(X, Y \given \Delta=1) \diff p(\theta) \\
    &=\ 0
    \label{eq:integrand_negative}
\end{align}

\paragraph{Proof of Equation \ref{eq:thm1p_ccl1}} First, we have:

\begin{align}
\SMI(X, Y) &= \EE[\PSMI(X, Y)] \\
&= \EE[\PSMI(X, Y) \given \Delta = 0]p(\Delta = 0) \\
& \quad + \EE[\PSMI(X, Y) \given \Delta = 1]p(\Delta = 1)
\end{align}

Using Equation \ref{eq:thm1p_ccl2} that we have proved, we obtain:

\begin{equation}
    \EE[\PSMI(X, Y) \given \Delta = 0] > \SMI(X, Y)
\end{equation}

As a result, it is sufficient to demonstrate Equation $\ref{eq:thm1_ccl1}$ with $\SMI(X, Y)$ instead of $\EE[\PSMI(X, Y) \given \Delta = 0]$. To do this, we will apply Theorem 1 in \citep{wongso_using_2023}. To do this, we search $(R_0, R_1, m_g, \nu) \in \RR_{+, *}^4$ such that $(X, Y)$ is $(R_1, R_2, m_g, \nu)$-SSM separated with respect to Definition 3 in \citep{wongso_using_2023}. Let $D = ||\mu_0 - \mu_1||$. Using $\mu_0$ and $\mu_1$ and the centers of the spheres, this means that $(R_0, R_1, m_g, \nu)$ should satisfy:

\begin{equation} \label{eq:ssm_separation}
\left \{ \begin{gathered}
    p(||X - \mu_0|| > R_0) = p(||X + \mu_1|| > R_1) = \nu \\
    R_0 + R_1 + m_g = D
\end{gathered} \right .
\end{equation}

There are many values of $(R_0, R_1, m_g, \nu)$ which satisfy these conditions. When applying Theorem 1 in \citep{wongso_using_2023}, these values give different lower bounds. Here is an algorithm to create a valid tuple $(R_0, R_1, m_g, \nu)$ given a hyperparameter $R \in ] 0, D/2[$.

\begin{enumerate}
    \item Let $S_0$ (resp. $S_1$) be the sphere of center $\mu_0$ (resp. $\mu_1$) and radius $R$.
    \item Let $\nu_0 = p(X \in S_0 \given Y=0)$ and $\nu_1 = p(X \in S_1 \given Y=1)$. Given the Gaussian assumptions we made in Equation \ref{eq:thm1p_hyp1}, we have $\nu_0, \nu_1 \in ]0, 1[$.
    \item Let $i \in \{0, 1\}$ and $j = i - 1$ such that $\nu_i \ge \nu_j$. We fix $R_i = R$ and $\nu = \nu_i$.
    \item We will now start with $R_j = R$ and decrease its value until Equation \ref{eq:ssm_separation} is satisfied. Because $X$ is a continuous random variable, the following function is continuous, decreasing, equal to $1$ when $t = 0$, and because $\nu_j \le \nu_i$, its value is $\le \nu$ for $t = 1$:
    \begin{equation}
        t \in [0, 1] \mapsto p(||X-\mu_j|| > t \cdot R \given Y = j)
    \end{equation}
    \item As a consequence, due to the intermediate values theorem, there exists $t_j$ in $]0, 1]$ such that $p(||X-\mu_j|| > t \cdot R \given Y = j) = \nu$. 
    \item We set $R_j = t \cdot R$ and $m_g = D - R_0 - R_1$. Because $R_0, R_1 \le R < D/2$, we have $m_g > 0$
    \item Now, we can apply Theorem 1 in \citep{wongso_using_2023} :
\end{enumerate}

\begin{equation}
\begin{gathered}
    \SMI(X, Y) \\
    > (1 - H(\nu, 1-\nu)) \times B_{\gamma(m_g, R_0, R_1)}\left( \frac{d-1}{2}, \frac{1}{2}\right)
\end{gathered}
\end{equation}

Where:

\begin{itemize}
    \item $H$ is the entropy function $H(p_1, p_2) = -p_1 \log p1 -p_2 \log p_2$. We can easily prove that $(1 - H(\nu, 1- \nu)$ is convex on $]0, 1[$ and that its minimal value is $> 0$.
    \item $\gamma(m_g, R_0, R_1) = \frac{m_g}{m_g + R_0 + R_1} \left( 2 - \frac{m_g}{m_g + R_0 + R_1} \right) = \frac{m_g}{D}\left(2 - \frac{m_g}{D} \right)\in ]0, 1[$
    \item $B$ is the incomplete beta function defined as follows. Because $\gamma(m_g, R_0, R_1) \in ]0, 1[$, it is clear that its value is always $>0$.
\end{itemize}

\begin{equation}
    B_\gamma(a, b) = \int_0^\gamma t^{a - 1}(1-t)^{b-1} \diff t
\end{equation}

This proves that $\SMI(X, Y) > 0$, which demonstrates Equation \ref{eq:thm1p_ccl2} and concludes the proof. \hfill $\qedsymbol$

\paragraph{Discussion on a better bound for Equation \ref{eq:thm1_ccl1}}

The proof above provides a constructive algorithm to obtain $(R_0, R_1, m_g, \nu) \in \RR_{+, *}^4$ such that $(X, Y)$ is $(R_1, R_2, m_g, \nu)$-SSM separated with respect to Definition 3 in \citep{wongso_using_2023}. Depending on the hyperparameter $R \in ]0, D/2[$, the bound is different. As a result, this hyperparameter can be optimized to find the better possible bound with this algorithm. We did not performed this optimization because it is not useful for the purpose of Theorem \ref{thm:psmi_outliers_proof}. Indeed, we only use this theorem to illustrate why we expect outliers in the hidden representations distribution to have a lower PSMI (see Section \ref{sec:preliminary}).
\section{Implementation details} \label{app:detail_predictors}

To help reproducing our results, we provide a GitHub repository containing the Python, Bash and Slurm code needed to deploy our experiments on a HPC cluster.\footnote{\safelink{https://github.com/orailix/predict_llm_memorization}{Project GitHub} } We also provide a PyPI package containing an automated estimator of PSMI that can be used in a wide range of scenarios.\footnote{\safelink{https://pypi.org/project/psmi/}{PSMI PyPI}} In this Section, we provide details on the implementation of our predictors $\hmu$ used in Algorithm \ref{alg:predict_memo}.

\textbf{For PSMI,} we use algorithm 1 in \citep{wongso_pointwise_2023}. We sample 2000 direction uniformly on the unit sphere. Indeed, we observed that the mutual information between random directions and the label has a mean of about $4.5 \cdot 10^{-3}$ and a standard deviation of about $5.5 \cdot 10^{-3}$. Thus, if we approximate these distributions by Gaussians, we get a margin at 95\% confidence interval of about $(1.96 \times 5.5 \cdot 10^{-3}) / \sqrt{2000} \simeq 2.4 \cdot 10^{-4}$. This is about 20 times smaller than the mean, so we consider that our metric is stable enough with 2000 estimators.

\textbf{For the loss,} we directly use the cross-entropy loss of the model for the last token before the label.

\textbf{For the logit gap,} we use the logits, which are the outputs of the fully-connected layer applied to the last token, before the softmax. The logit gap is the difference between the logit of the correct prediction and the maximum logit of an incorrect prediction. 

\textbf{For Mahalanobis distance \citep{mahalanobis_generalized_1936},} we compute the distance of the hidden representation of a training sample to the distribution of hidden representation of the other training samples. To reduce computational costs, we first project every hidden states using a Principal Component Analysis (PCA) with a target dimension of 500. This metric was suggested by 

\textbf{For the baseline,} see Section \ref{sec:eval_protocol}.

\section{Remarks on the baseline} \label{sec:discussion_baseline}

As mentioned in Section \ref{sec:eval_protocol}, the baseline used for comparison is an adaptation of the method proposed by \citet{biderman_emergent_2023}. Below, we demonstrate that the performance we measure for the baseline is consistent with the results reported in the original paper. Consequently, the poor performance of the baseline in our evaluations is not due to its adaptation to classification, but rather to a structural limitation of the approach of the baseline within this threat model.

\citet{biderman_emergent_2023} use precision and recall to evaluate their prediction. On the opposite, we use TPR and FPR. Recall is equal to the TPR. However, precision is not equal to the FPR, becauseit depends on the prevalence of memorized samples in the dataset. 

\paragraph{Computing the FPR of \citet{biderman_emergent_2023}}

Let TP, FP, P and N be the True Positive, False Positive, Positive and Negative samples. We define:

\begin{equation}
\begin{gathered}
    \TPR = \frac{\TP}{\PP} \quad \text{\textit{("true positive rate" or "recall")}} \\
    \FPR = \frac{\FP}{\NN} \quad \text{\textit{("false positive rate")}} \\
    \PVL = \frac{\PP}{\PP + \NN} \quad \text{\textit{("prevalence")}} \\
    \PRC = \frac{\TP}{\TP + \FP} \quad \text{\textit{("precision")}} \\
\end{gathered}
\end{equation}

By definition, we have:

\begin{equation}
    \FP = \frac{\TP}{\PRC} - \TP \quad \text{and} \quad
    \NN = \frac{\PP}{\PVL} - \PP
\end{equation}

Consequently, we have:

\begin{align}
    \FPR &= \frac{\frac{\TP}{\PRC} - \TP}{\frac{\PP}{\PVL} - \PP} \\
    &= \frac{\TP}{\PP} \times \frac{\frac{1}{\PRC} - 1}{\frac{1}{\PVL} - 1} \\
    &= \TPR \times \frac{\PVL (1 - \PRC)}{\PRC (1 - \PVL)}
\end{align}

In Table~2a~\citep{biderman_emergent_2023}, they report PRC=0.919 and TPR=0.513 for a Pythia 6.9B model \citep{biderman_pythia_2023} trained for $23 \cdot 10^6 / 146 \cdot 10^6 \simeq 15.7\%$ of the training steps.  This model has a similar size and architecture to the ones we evaluate. Given the prevalence of $\PVL \simeq 1.1\%$ reported in their paper, this corresponds to a $\FPR \simeq 0.05\%$.

\paragraph{Discussion}

\begin{figure*}[ht!]
    \centering
    \includegraphics[width=\textwidth]{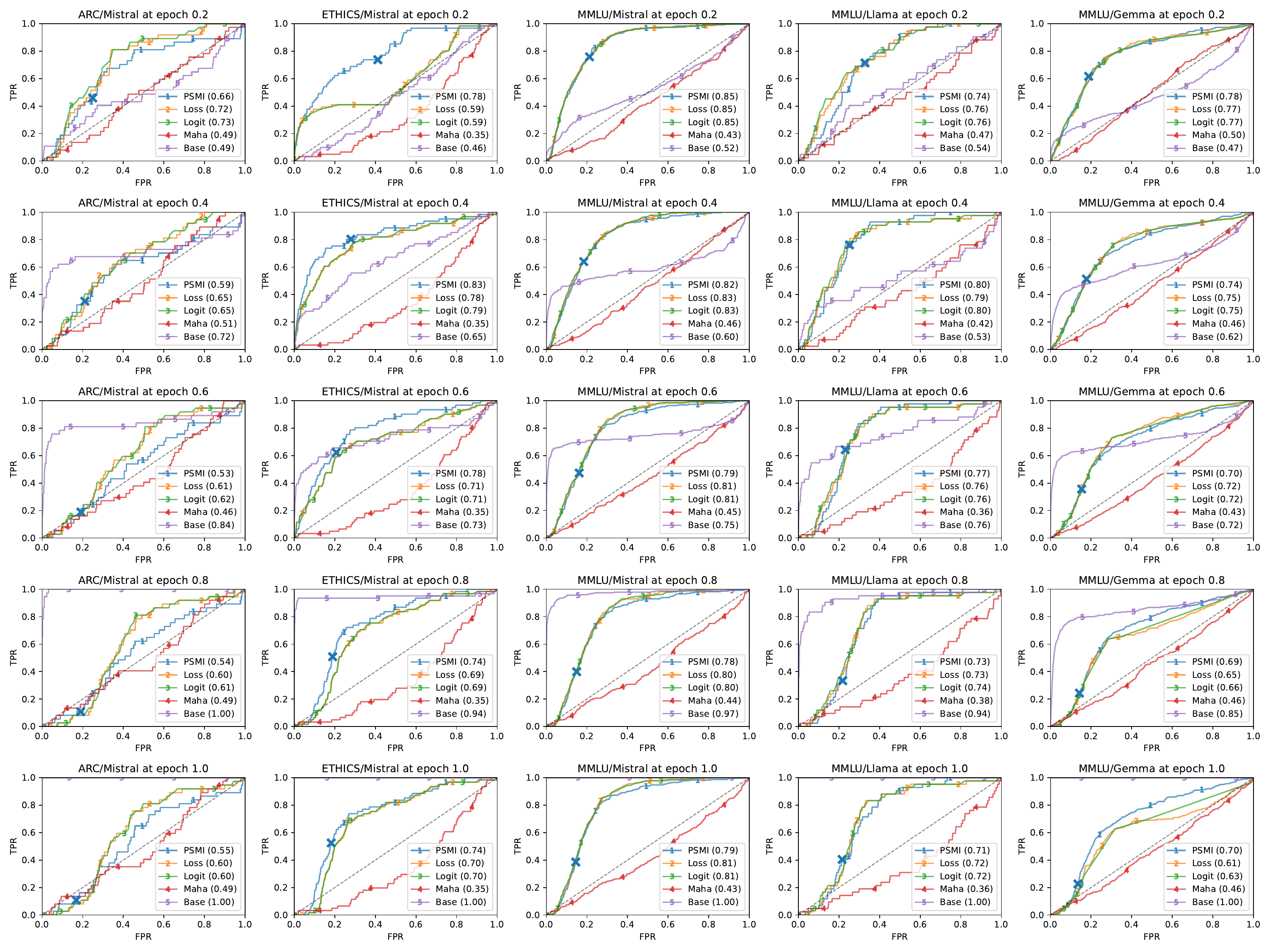}
        \caption{TPR/FPR trade-offs for each setting at every checkpoint. The number in parentheses corresponds to the AUC. The blue cross indicates the result using the recommended hyperparameters for practitioners in Table \ref{tab:hpar_values}. The AUC of the baseline ("Base") converges to 1.0 at epoch 1.0 because, at that stage, the baseline is the same as what we are trying to predict, by definition. We remind that practitioners within our threat model have the resources to compute neither the baseline nor ground truth memorization.}
    \label{fig:appendix_mu}
\end{figure*}

The very low FPR indicates that their method makes very few mistakes: a sample memorized by the partially trained model is extremely unlikely to be \textit{forgotten} by the fully trained model. We also observe behavior with the baseline, as evidenced by the steep slope of the baseline's ROC curve near the origin (see Figures \ref{fig:pred_perfs} and \ref{fig:appendix_mu}). In the experiments of Section \ref{sec:experiments}, the baseline never achieves a TPR of 51.3\% with a TPR as low as 0.05\%. Indeed, we predict memorization \textit{before} it occurs, which structurally limits the performance of the approach of \citet{biderman_emergent_2023}, as it relies on memorization in the partially trained model. If we wait longer and make the prediction at epoch 0.6 out of 1, the baseline achieves values close to TPR=51.3\% / FPR=0.05\% (see third line in Figure \ref{fig:appendix_mu}). However, waiting so long before making the prediction contradicts our threat model and is not applicable for practitioners wishing to implement mitigation techniques (see Section \ref{sec:problem_setting}). 

Conversely, the approach we propose overcomes the limitations of the baseline by predicting memorization before it occurs. It is the first to achieve both a low FPR and a high TPR within this threat model.
\section{Additional experiments} \label{app:additional_plots}

For brevity, each experiment in Section 3 is presented for only one dataset/model. However, all experiments were run on all datasets and are presented in this section.

\paragraph{Special case of ARC/Mistral} Table \ref{tab:loss_decrease} presents the decrease of the median training loss relative to epoch 0 throughout training. We saved models every 0.2 epoch to analyze them and measure their performance. We observe that for ARC/Mistral, the median training loss has decreased by 93.724\% at epoch 0.2, which is close to 95\%. Conversely, by epoch 0.4, the median training loss has decreased by significantly more than 95\%. This is why, for that setting, we predict memorization at epoch 0.2, the checkpoint where the decrease is closest to 95\%. For the other settings, as indicated in Section \ref{sec:experiments}, we predict memorization at the first checkpoint where the median training loss has decreased by at least 95\%.

\begin{table}[h!]
\centering
\begin{adjustbox}{width=\columnwidth}
\begin{tabular}{lcccccc}
\toprule
 & Epoch 0 & Epoch 0.2 & Epoch 0.4 & Epoch 0.6 & Epoch 0.8 & Epoch 1 \\
\midrule
ARC/Mistral & 0.000\% & 93.724\% & 98.015\% & 99.397\% & 98.907\% & 99.222\% \\
ETHICS/Mistral & 0.000\% & 86.036\% & 95.198\% & 98.985\% & 99.665\% & 99.739\% \\
MMLU/Mistral & 0.000\% & 98.967\% & 99.776\% & 99.916\% & 99.916\% & 99.895\% \\
MMLU/Llama & 0.000\% & 91.674\% & 98.186\% & 99.329\% & 99.336\% & 99.267\% \\
MMLU/Gemma & 0.000\% & 99.543\% & 99.606\% & 99.855\% & 99.980\% & 99.979\% \\
\bottomrule
\end{tabular}
\end{adjustbox}
\caption{Decrease in the median training loss relative to epoch~0 throughout training.}
\label{tab:loss_decrease}
\end{table}
\begin{figure*}[ht!]
    \centering
    \includegraphics[width=\textwidth]{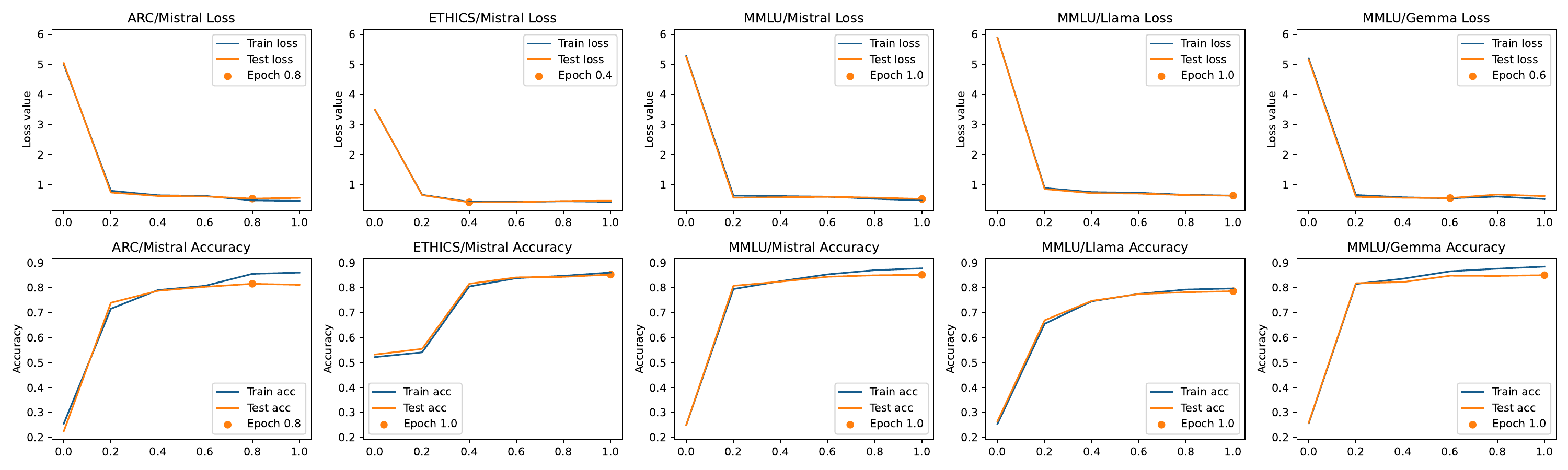}
    \includegraphics[width=\textwidth]{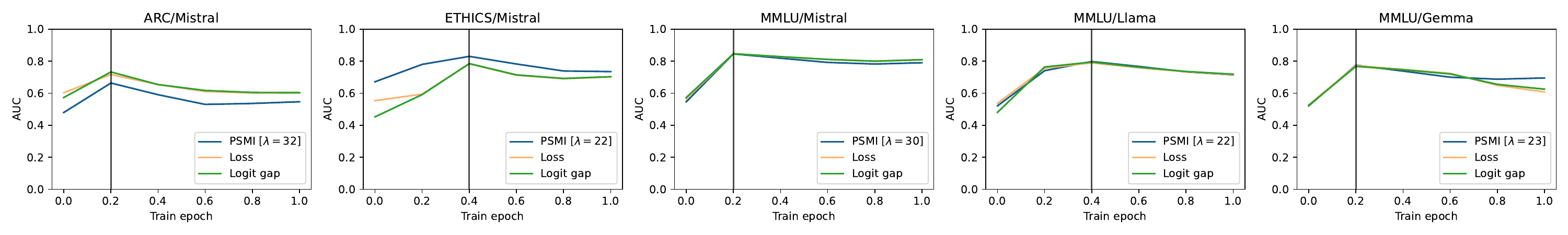}
    \includegraphics[width=\textwidth]{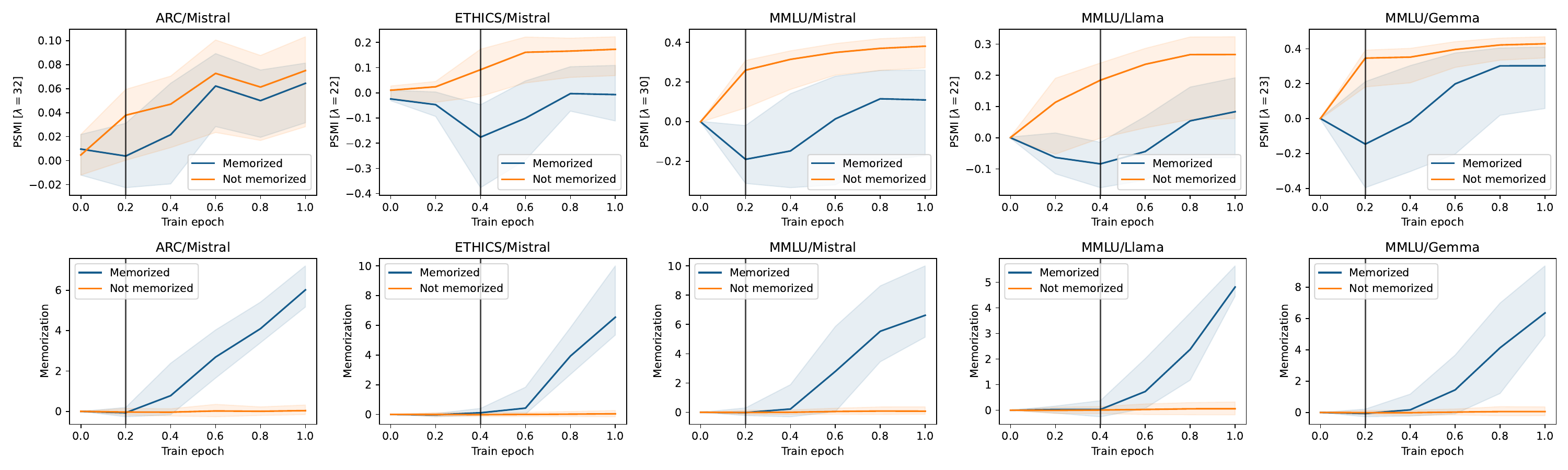}
    \caption{Additional results related to the dynamics of training and the appropriate moment to interrupt training. \textbf{First row:} Training loss, testing loss, and epoch of the best testing loss. \textbf{Second row:} Training accuracy, testing accuracy, and epoch of the best testing accuracy. \textbf{Third row, Fourth row and Fifth row:} Similar to Figure \ref{fig:ablation_rho}.}
    \label{fig:appendix_rho}
\end{figure*}
\begin{figure*}[ht!]
    \centering
    \includegraphics[width=\textwidth]{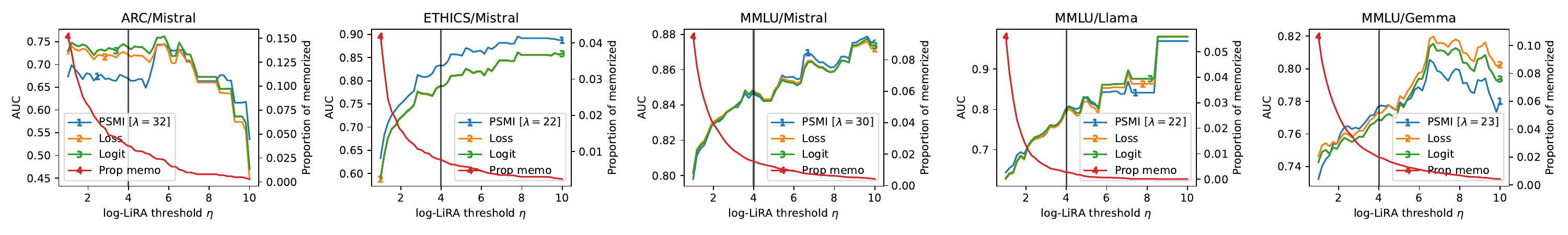}
        \caption{Similar to Figure \ref{fig:ablation_lambda_and_eta} (Right): impact of the threshold $\heta$.}
        \label{fig:appendix_eta}
\end{figure*}
\begin{figure*}[ht!]
    \centering
    \includegraphics[width=\textwidth]{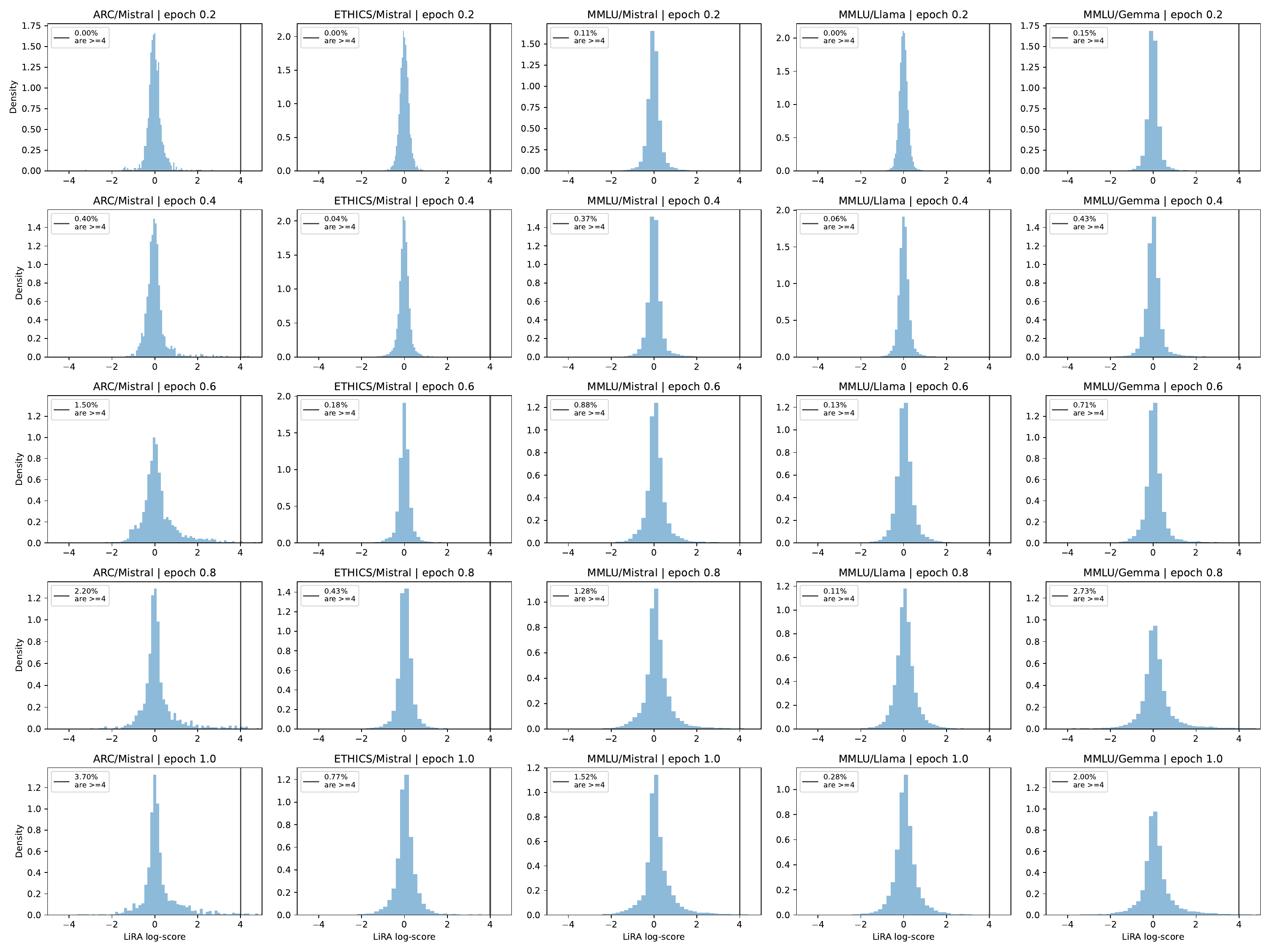}
        \caption{Histograms of memorization throughout training. The legend displays the proportion of samples with log-LiRA $\ge 4$.}
    \label{fig:appendix_memo_histogram}
\end{figure*}
\begin{figure*}[t!]
    \centering
    \includegraphics[width=0.27\textwidth]{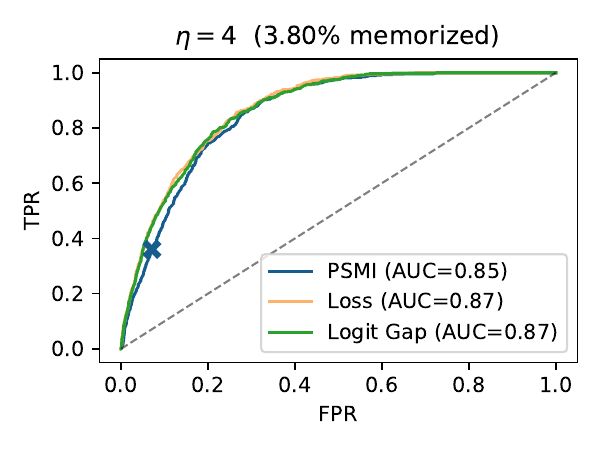}
    \includegraphics[width=0.27\textwidth]{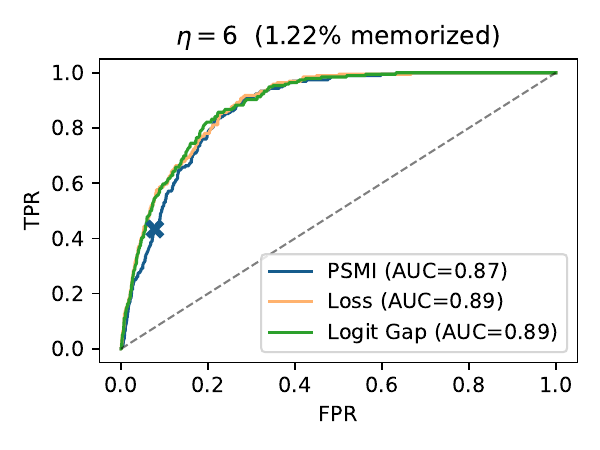}
    \includegraphics[width=0.27\textwidth]{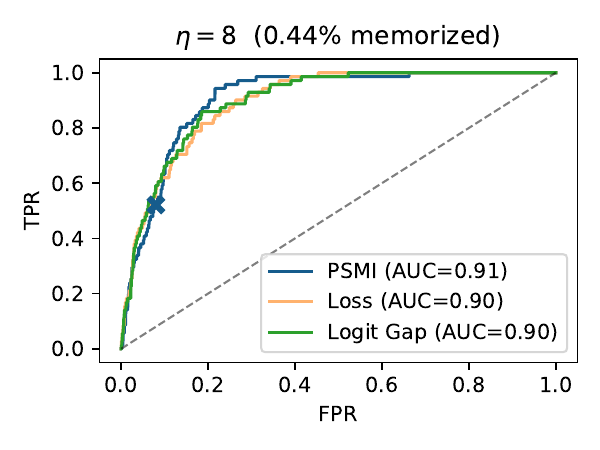}
    \includegraphics[width=0.27\textwidth]{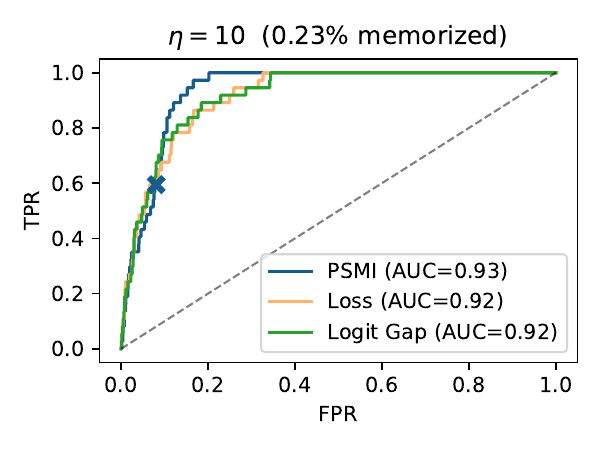}
    \includegraphics[width=0.27\textwidth]{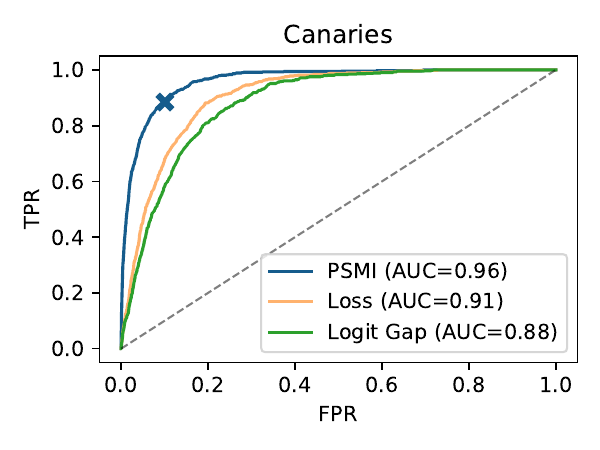}
    \includegraphics[width=0.27\textwidth]{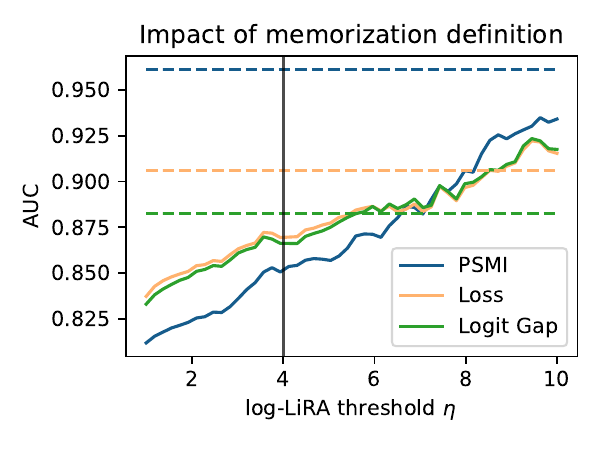}
    \caption{Plots 1-4: Similar to Figure \ref{fig:ablation_cifar} (Left) with different values for $\heta$. Plots 5-6: Similar to Figure \ref{fig:ablation_cifar} (Center and Right).}
    \label{fig:appendix_cifar}
\end{figure*}

\end{document}